\documentclass[aps,pra,twocolumn,superscriptaddress,floatfix]{revtex4-1}
\usepackage{ifpdf}
 \newif\ifpdfq

\ifx\pdfoutput\undefined
   \pdffalse
\else
   \pdfoutput=1
   \pdftrue
\fi
\ifpdf
   \usepackage{graphicx}
   \usepackage{epstopdf}
   \usepackage{color}
\else
   \usepackage{graphicx}
\fi

\usepackage{srctex}
\usepackage{hyperref}
\usepackage{bm}

\begin{document}


\title{Theoretical and experimental study of \\ high-pressure synthesized B20-type compounds Mn$_{1-x}$(Co,Rh)$_x$Ge}

\date{\today}

\author{N.~M.~Chtchelkatchev}
\affiliation{Landau Institute for Theoretical Physics, Russian Academy of Sciences,
142432 Chernogolovka, Moscow Region, Russia}
\affiliation{Vereshchagin Institute for High Pressure Physics, Russian Academy of Sciences,
108840 Troitsk, Moscow, Russia}
\affiliation{Moscow Institute of Physics and Technology, 141700 Dolgoprudny, Moscow Region, Russia}
\affiliation{Ural Federal University, 620002, 19 Mira str., Ekaterinburg, Russia}

\author{M.~V.~Magnitskaya}
\affiliation{Vereshchagin Institute for High Pressure Physics, Russian Academy of Sciences,
108840 Troitsk, Moscow, Russia}
\affiliation{Lebedev Physical Institute, Russian Academy of Sciences, 119991 Moscow, Russia}
\affiliation{Landau Institute for Theoretical Physics, Russian Academy of Sciences,
142432 Chernogolovka, Moscow Region, Russia}

\author{V.~A.~Sidorov}
\affiliation{Vereshchagin Institute for High Pressure Physics, Russian Academy of Sciences,
108840 Troitsk, Moscow, Russia}

\author{L.~N.~Fomicheva}
\affiliation{Vereshchagin Institute for High Pressure Physics, Russian Academy of Sciences,
108840 Troitsk, Moscow, Russia}

\author{A.~E.~Petrova}
\affiliation{Vereshchagin Institute for High Pressure Physics, Russian Academy of Sciences,
108840 Troitsk, Moscow, Russia}

\author{A.~V.~Tsvyashchenko}
\affiliation{Vereshchagin Institute for High Pressure Physics, Russian Academy of Sciences,
108840 Troitsk, Moscow, Russia}

\begin{abstract}

The search and exploration of new materials not found in nature is one of modern trends in pure and applied chemistry. In the present work, we report on experimental and \textit{ab initio} density-functional study of the high-pressure-synthesized series of compounds Mn$_{1-x}$(Co,Rh)$_x$Ge. These high-pressure phases remain metastable at normal conditions, therewith they preserve their inherent noncentrosymmetric B20-type structure and chiral magnetism. Of particular interest in these two isovalent systems is the comparative analysis of the effect of $3d$ (Co) and $4d$ (Rh) substitution for Mn, since the $3d$ orbitals are characterized by higher localization and electron interaction than the $4d$ orbitals. The behavior of Mn$_{1-x}$(Co,Rh)$_x$Ge systems is traced as the concentration changes in the range $0 \leq x \leq 1$. We applied a sensitive experimental and theoretical technique which allowed to refine the shape of the temperature dependencies of magnetic susceptibility $\chi(T)$ and thereby provide a new and detailed magnetic phase diagram of Mn$_{1-x}$Co$_x$Ge. It is shown that both systems exhibit a helical magnetic ordering that very strongly depends on the composition $x$. However, the phase diagram of Mn$_{1-x}$Co$_x$Ge differs from that of Mn$_{1-x}$Rh$_x$Ge in that it is characterized by coexistence of two helices in particular regions of concentrations and temperatures.

\end{abstract}


\maketitle

\section{Introduction}

\begin{figure*}
  \centering
  \includegraphics[width=\columnwidth]{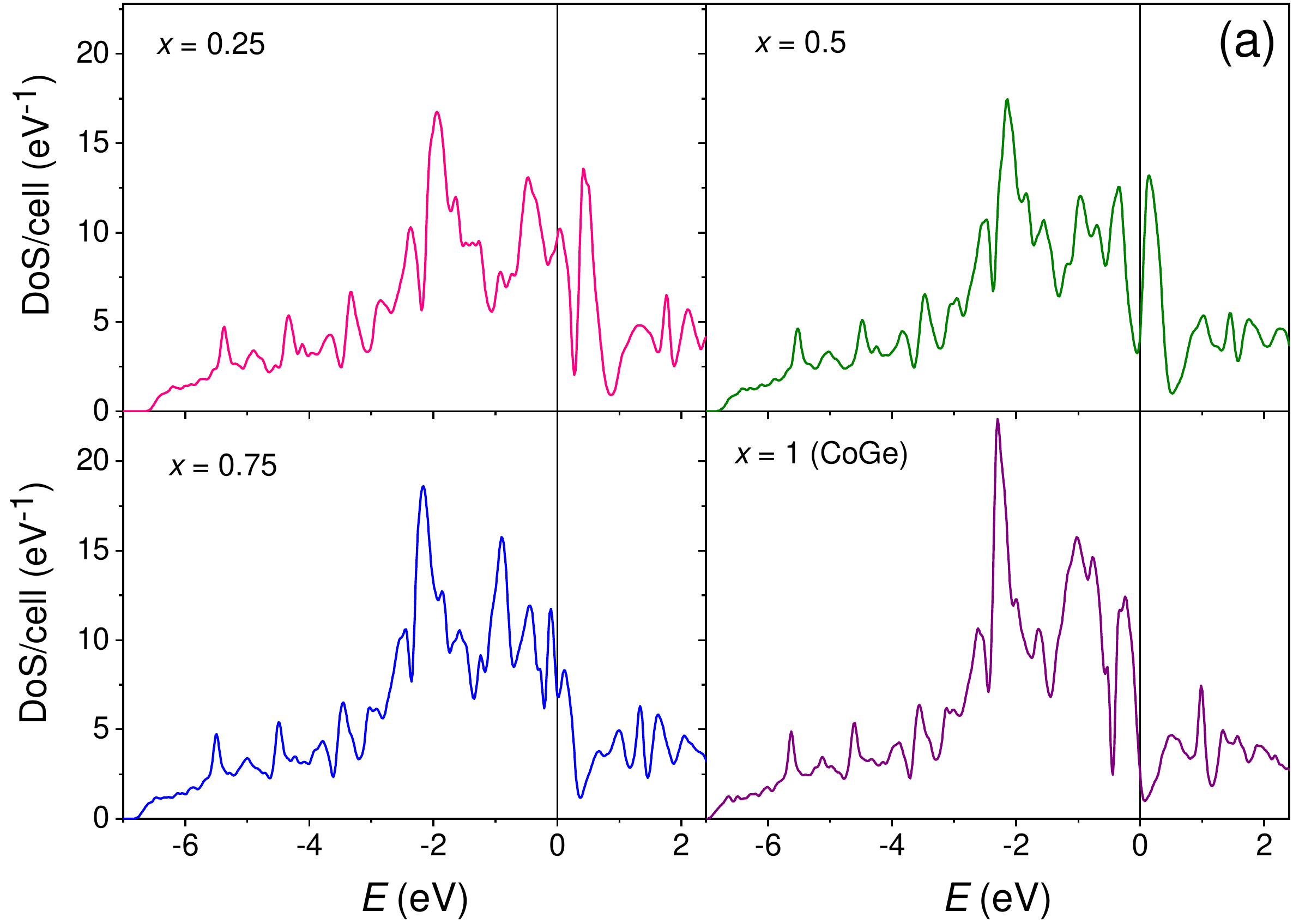} \includegraphics[width=\columnwidth]{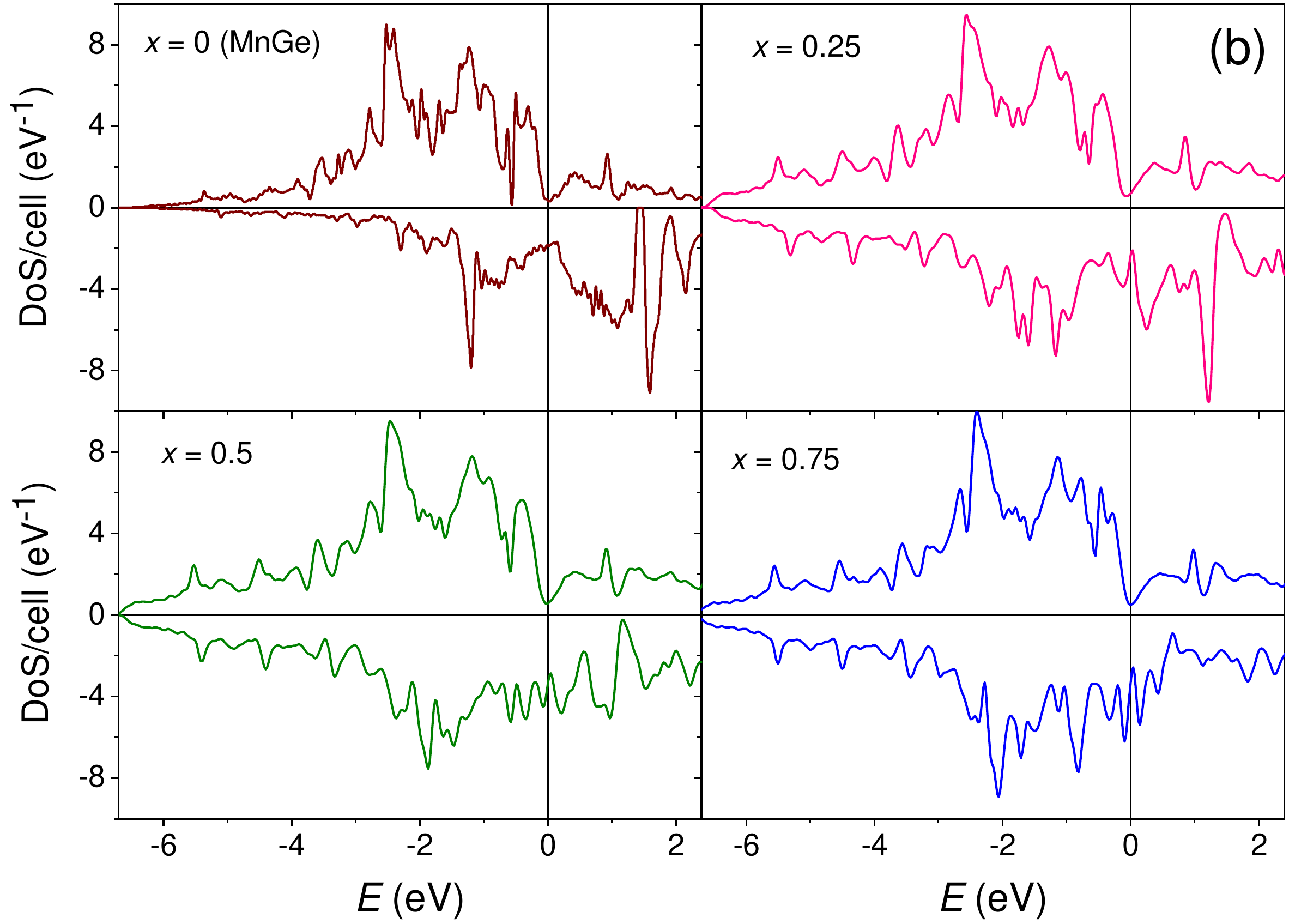}
  \caption{(Color online) The results of non-spin-polarized (a) and spin-polarized (b) calculations of the density of states for Mn$_{1-x}$Co$_x$Ge ( $x = 0.25, 0.5,$ and 0.75). The DOS of magnetic MnGe and nonmagnetic CoGe are also shown. The Co content $x$ is indicated in corresponding panels. The DOS for the spin up (down) states in Fig.~(b) is counted positive (negative). The Fermi level  ($E_{\mathrm{F}}$) is set to zero and marked by the vertical line.}\label{fig:RhCo-dos}
\end{figure*}

The design, fabrication and study of novel compounds not found in nature, with an emphasis on interrelation between their electronic structure and physico-chemical properties, provide a promising basis for the development of technologically important materials~\cite{Tretyakov2004PAC,Manyala2004,Martin2014PAC,DiTusa2014,Dmitrienko2014Nature,Koretsune2015SR,Wu2018SR}. Examples of such systems are transition-metal (TM) monogermanides with a cubic B20-type crystal structure, which are attractive for researchers in view of their exotic properties and hence, a variety of possible applications. Most of these materials have been obtained by the direct high-pressure synthesis from melted constituents and found to remain metastable under normal conditions~\cite{sidorov2018}.

These systems are less studied than their isostructural analogues — B20-type TM monosilicides, because the latters are normal-pressure phases and as such, easier to grow and explore. However, the TM monogermanides generally exhibit a much wider spectrum of exotic magnetic phenomena than monosilicides. For example, in contrast to a number of TM monogermanides that possess chiral long-period magnetic structures at relatively high temperatures, the monosilicide systems that form continuous series of compounds in the composition range $0 \leq x \leq 1$, are magnetic only in a limited interval of concentrations (Mn$_{1-x}$Fe$_x$Si) or temperatures (Fe$_{1-x}$Co$_x$Si)~\cite{Manyala2004}. This circumstance makes TM monogermanides good candidates to systematically study the effect of substituting one TM constituent by another, with the latter not necessarily being 3d magnetic element.


A nontrivial chiral magnetic ordering found in some B20-type TM monogermanides is determined by the competition between magnetic exchange and spin-orbit  interactions, and the interplay of structural and magnetic chiralities. There is RKKY interactions between magnetic atoms in the system, while the noncentrosymmetric nature of the B20 lattice induces the Dzyaloshinskii--Moriya interaction (DMI) where its sign is correlated with the B20 structural chirality~\cite{Jensen1980JPhys,jensen1991rare,grigoriev2013}. Experimental studies show that the period, sign, and number of occurring magnetic helices strongly depend also on the chemical composition, i.e. symmetry, the cell volume, the lengths and angles of chemical bonds, the concentration and type of dopant (3d or 4d, donor or acceptor, magnetic or nonmagnetic). Thus, the choice of transition-metal constituents of monogermanides can strongly affect the sign of DMI~\cite{Siegfried2015PRB}, which is important for the search of candidate spintronics-related materials.

In the present work, we report on experimental and theoretical study of the high-pressure-synthesized chiral magnets Mn$_{1-x}$Co$_x$Ge. The behavior of this system is traced as cobalt concentration changes in the range $0 \leq x \leq 1$. It should be noted that compositions belonging to the Mn-rich side ($x \leq 0.5$) have been studied in~\cite{martin2017} using the small-angle neutron scattering (SANS) and neutron powder diffraction techniques. More recently, the characterization of magnetic structure in Mn$_{1-x}$Co$_x$Ge at $0 \leq x \leq 0.9$ has been performed using the SQUID magnetometer and SANS measurements~\cite{altynbaev2018}. Here, we applied a more sensitive experimental technique, which allowed us to refine the shape of the $\chi(T)$ dependencies and thereby provide a new magnetic phase diagram of Mn$_{1-x}$Co$_x$Ge. Our experimental data are theoretically analyzed on the basis of  \textit{ab initio} density-functional calculations of Mn$_{1-x}$Co$_x$Ge at various concentrations $x$.

Then we compare the theoretical and experimental results for two series of pseudobinary alloys, Mn$_{1-x}$Rh$_x$Ge~\cite{sidorov2018} and Mn$_{1-x}$Co$_x$Ge, to study the effect of 3d- and 4d-doping on the evolution of structural, electronic, magnetic, and transport properties. At high Co/Rh concentrations $x \geq 0.5$, helimagnetic structures with long periods are observed: $\sim 550$~{\AA} for Mn$_{1-x}$Rh$_x$Ge and $\sim 380$~{\AA} for Mn$_{1-x}$Co$_x$Ge~\cite{martin2017}, which are more than one order of magnitude larger than in pure MnGe and the Mn-rich side ($x \leq 0.25$) of the Mn$_{1-x}$(Co,Rh)$_x$Ge series~\cite{martin2017,altynbaev2018}.

The organization of the paper is as follows: In Sec.~\ref{Sec2} technical details of experimental measurements are given and theoretical calculations are described. Structural properties are considered in Sec.~\ref{Sec3}. Sec.~\ref{Sec4} presents the results of magnetic susceptibility measurements and corresponding data processing. Section \ref{Sec5} is devoted to the comparison of magnetic and transport properties and their discussion for both Mn$_{1-x}$Co$_x$Ge and Mn$_{1-x}$Rh$_x$Ge. Finally, in Sec.~\ref{Sec6} we sum up the results.

\section{Methods\label{Sec2}}

\subsection{Experimental technique}

 Polycrystalline samples of the B20-type compounds Mn$_{1-x}$Co$_x$Ge ($0 \leq x \leq 1$) were synthesized under pressure $P$ = 8~GPa by melting Mn, Co and Ge in the toroidal high-pressure cell~\cite{Khvostantsev2004}. The obtained high-pressure phases remain metastable at normal conditions. The crystal structure was analyzed by X-ray diffraction (XRD) at room temperature and normal pressure using the diffractometer STOE IPDS-II (Mo-K$\alpha$). For details of the experimental procedure see paper~\cite{sidorov2018} and references therein.

The electrical resistivity was measured on bulk polycrystalline samples with the help of lock-in detection technique (SR830 lock-in amplifier and SR554 preamplifier) in the temperature range 4 to 300~K. In this case, four Pt electrodes of 25 $\mu$m in diameter were spot-welded to the sample.

To obtain the temperatures of magnetic ordering in the entire range of concentrations $x$, the magnetic ac-susceptibilities ($\chi$) were measured with SR830 lock-in amplifier and home-made coil system in the range 4 to 300~K. Signal of SR830 was normalized to the sample mass, and all measurements were performed at fixed frequency and excitation field, so the susceptibilities in arbitrary units may be compared for different samples on the absolute scale. The excitation field was of about 1 Oe that is much smaller than in SQUID measurements (1000 Oe)~\cite{altynbaev2018}. Small excitation field makes it possible to resolve double-peak features of $\chi(T)$ dependencies (see Sec.~\ref{Sec4}) that are smeared out in higher fields.

\subsection{Density functional calculations}

Our \textit{ab initio} computations are based on the density functional theory (DFT). We used the first-principles pseudopotential method as implemented in the Quantum Espresso package~\cite{Giannozzi2009}, with the exchange-correlation functional taken within generalized-gradient approximation (GGA) by Perdew-Burke-Ernzerhof (PBE)~\cite{Perdew1996}. We employed the projected-augmented-wave (PAW) type scalar-relativistic pseudopotentials from the Quantum Espresso database, with the valence electron configurations of $3s^2p^6d^54s^2$, $3s^2p^6d^74s^2$, and $3d^{10}4s^2p^2$ for Mn, Co, and Ge, respectively. The integration over the Brillouin zone (BZ) for the electron density of states computation was performed on a uniform grid of $24 \times 24 \times 24$ $k$-points. For $x = 1/4, 1/2$, and 3/4, the due number of equivalent Mn atoms in the cubic B20 unit cell were replaced by Co atoms. The plane wave cutoff of 100 Ry was chosen, which gives the total energy convergence better than $10^{-8}$~Ry. For each composition, the equilibrium value of system`s lattice constant $a_0$ was defined as the one corresponding to zero pressure. The geometry relaxation was performed until the residual atomic forces were converged down to 3~meV/\AA. The optimized internal atomic positions for MnGe are $u_{\mathrm{Mn}} = 0.135$ and $u_{\mathrm{Ge}} = 0.843$, while for CoGe $u_{\mathrm{Rh}} = 0.137$ and $u_{\mathrm{Ge}} = 0.840$ (experimental values are 0.128 and 0.834).

We made \textit{ab initio} density-functional calculations of the compounds Mn$_{1-x}$Co$_x$Ge at $0 \leq x \leq 1$, with and without taking into account the spin polarization. Our spin-polarized calculations were done using a simple model of collinear ferromagnetism. With increasing Co concentrations, a change occurs from a relatively short-period helix (SPH) to a long-period ($\sim 550$~{\AA}) helix (LPH). The simple ferromagnetic alignment is a reasonable approximation,  because even the short period ($\sim 30$~{\AA} as in pure MnGe) of spiral magnetic structure is significantly longer than the unit-cell size $\sim 4.5$~{\AA}. Both nonmagnetic and magnetic solutions are obtained at all concentrations, except for pure CoGe, which is nonmagnetic.

The calculated electron density of states (DOS) of Mn$_{1-x}$Co$_x$Ge for the magnetic and nonmagnetic phases is presented in FIG.~\ref{fig:RhCo-dos}. Our calculations show that over the entire energy range, the DOS, $N(E)$, of Mn$_{1-x}$Co$_x$Ge is contributed mostly by transition-metal $3d$-states hybridized with germanium $p$-states with a dominating contribution from the former. As could be expected, the DOS of Mn$_{1-x}$Co$_x$Ge is very similar to that of isovalent Mn$_{1-x}$Rh$_x$Ge calculated in our paper~\cite{sidorov2018}.

\section{Crystal structure of Mn$_{1-x}$(Co,Rh)$_x$Ge\label{Sec3}}

\subsection{Powder X-ray diffraction}

\begin{figure}[t]
  \centering
  \includegraphics[width=\columnwidth]{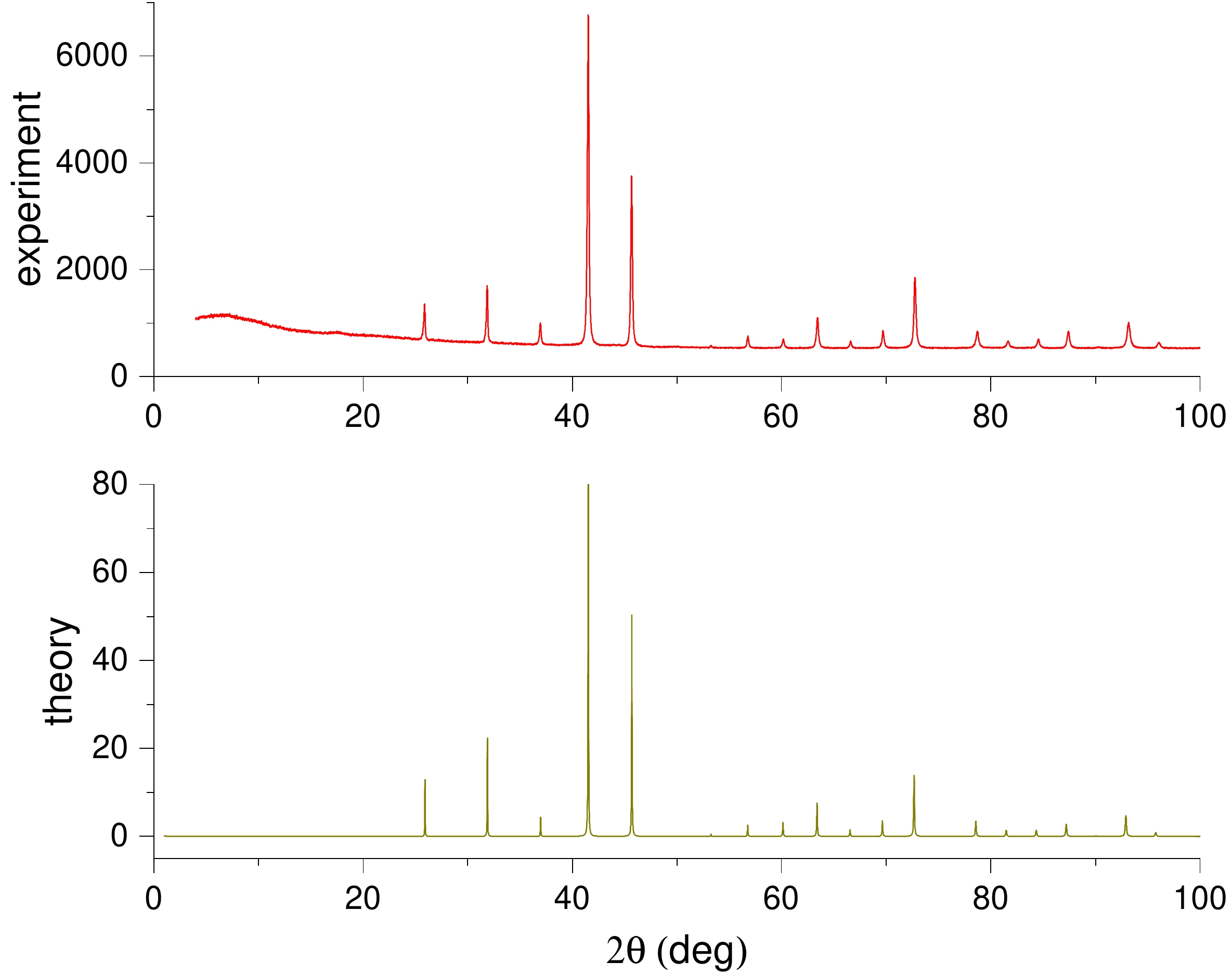}
  \caption{(Color online) Experimental (top panel) and simulated (bottom panel) X-ray powder-diffraction data for RhGe. Theoretical graph in bottom panel is obtained using the VESTA software that processes DFT-calculated data. We used as input to VESTA the B20-RhGe lattice parameters fully relaxed at normal pressure using the Quantum Espresso package.}\label{fig:RhGediffraction}
\end{figure}

\begin{figure}
  \centering
  \includegraphics[width=\columnwidth]{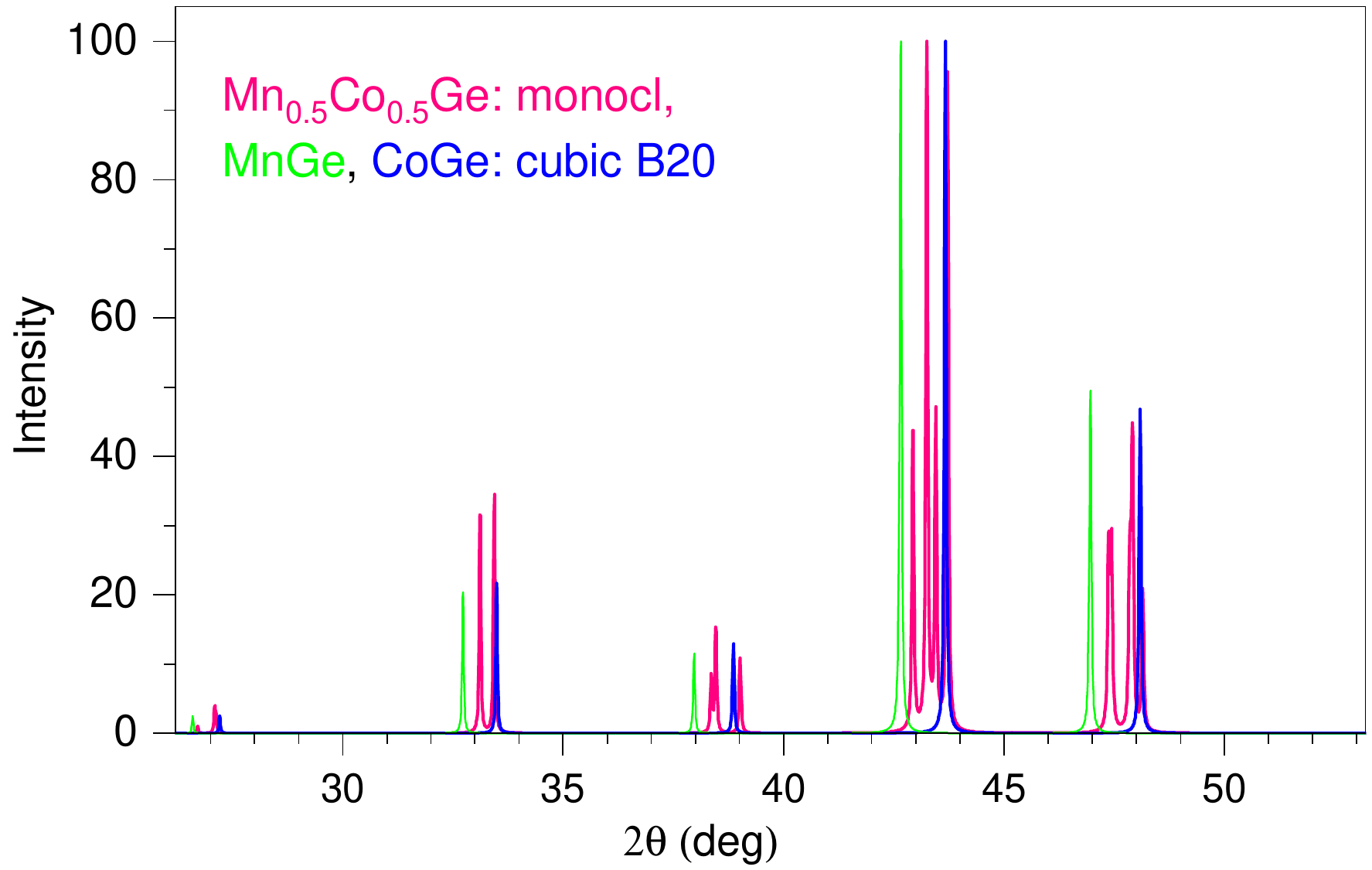}
  \caption{(Color online) The X-ray powder diffraction patterns calculated for Mn$_{0.5}$Co$_{0.5}$Ge, and B20-type binaries MnGe and CoGe. As follows, the diffraction peaks of Mn$_{0.5}$Co$_{0.5}$Ge split due to symmetry-breaking distortions, but they do not deviate far from the positions corresponding to the B20 lattice.}\label{fig:diffraction}
\end{figure}

High-pressure-synthesized binary monogermanides composed of 3d and Ge atoms crystallize in the B20 structure as confirmed by powder XRD measurements. For example, Fig.~\ref{fig:RhGediffraction} shows our simulated powder-diffraction data for RhGe in comparison with the experiment~\cite{Tsvyashchenko2016}. The theoretical XRD pattern (Fig.~\ref{fig:RhGediffraction}, bottom panel) is generated using the VESTA (Visualization for Electronic and STructural Analysis) software package~\cite{VESTA} that processes results of DFT computations. Here, our calculated parameters of the B20-RhGe structure were fully relaxed at $P=0$ with Quantum Espresso and then used as input data for VESTA. It is worth noting that the theoretical and experimental diffraction graphs are very similar to each other in positions and relative height of the peaks (one should bear in mind that the experimental XRD has been measured on polycrystalline samples).

When a particular compound contains both the Mn and Co/Rh atoms, a symmetry-breaking distortion of the B20 lattice takes place. Hence, at $x = 0.25, 0.5$ and 0.75, the lattice can be only approximately considered as B20-type. It is clear that the compounds Mn$_{0.5}$(Co,Rh)$_{0.5}$Ge possess maximal lattice distortion among the other compositions considered. However, the distortion from B20 is not large. To illustrate that we show in Fig.~\ref{fig:diffraction} the XRD pattern calculated for Mn$_{0.5}$Co$_{0.5}$Ge, and pure B20 binaries MnGe and CoGe. As follows, the diffraction peaks of Mn$_{0.5}$(Co,Rh)$_{0.5}$Ge split due to symmetry-breaking distortions, but they do not deviate far from the positions corresponding to the B20 lattice. This is also seen in experimental diffraction patterns of Mn$_{0.5}$(Co,Rh)$_{0.5}$Ge.

\subsection{Vegard's and non-Vegard's behavior}

\begin{figure}
  \centering
  \includegraphics[width=\columnwidth]{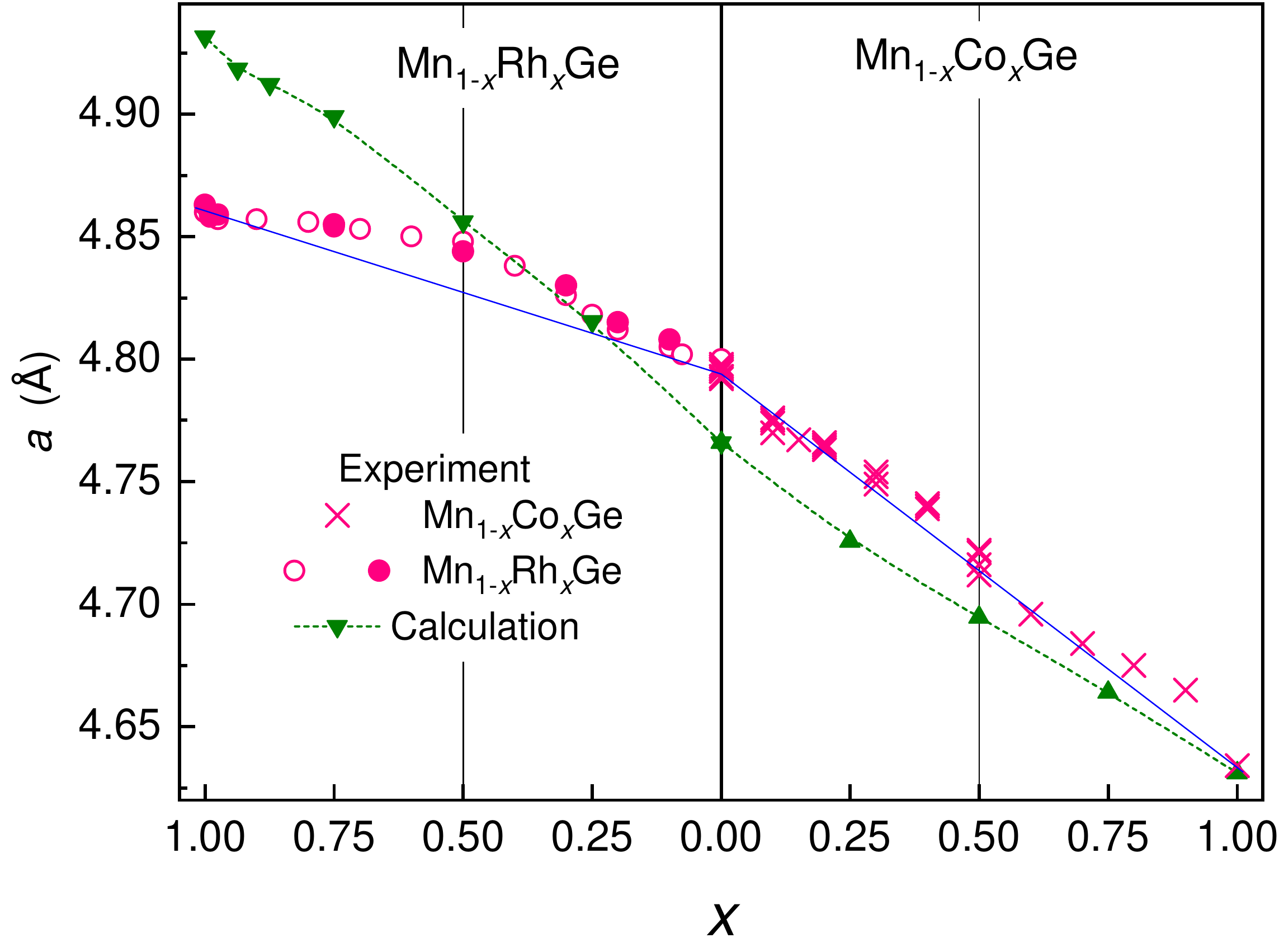}
  \caption{(Color online) The measured and calculated concentration dependence of the lattice parameter $a$ for Mn$_{1-x}$Co$_x$Ge in comparison with Mn$_{1-x}$Rh$_x$Ge~\cite{sidorov2018}.}\label{fig:a-x}
\end{figure}

The XRD measurements of the structure of Mn$_{1-x}$Co$_x$Ge were made at many $x$ values in the range of $0 \leq x \leq 1$. A comparison of variation in the lattice parameter $a$ with 3d-Co/4d-Rh concentration $x$ is presented in Fig.~\ref{fig:a-x}. The results for Mn$_{1-x}$Rh$_x$Ge are taken from our paper~\cite{sidorov2018}. As is seen in the figure, the cubic lattice parameter $a$ of Mn$_{1-x}$Co$_x$Ge decreases with decreasing Co content, because the atomic radius of cobalt is a few percent smaller than that of manganese. In case of Mn$_{1-x}$Rh$_x$Ge the situation is reversed.

As is well known, the lattice parameter $a$ of a solid solution of two constituents with the same crystal structure [here, MnGe and (Co,Rh)Ge] obeys Vegard's law (blue straight llines), i.e. $a$ changes almost linearly with concentration $x$ of substituting atom. However, for Mn$_{1-x}$Rh$_x$Ge there is a strong positive deviation of the experimentally measured $a(x)$ from linear Vegard's law (left panel in Fig.~\ref{fig:a-x}), while $\mathrm{Mn}_{1–x}\mathrm{Co}_x\mathrm{Ge}$ demonstrates much smaller deviation if any (right panel in Fig.~\ref{fig:a-x}). Similar practically linear $a(x)$ curve has been already observed for Mn$_{1-x}$Co$_x$Ge in papers~\cite{valkovskiy2016,altynbaev2018}. The linear dependence $a(x)$ has been also reported for other solid solutions of B20-type 3d-metal monogermanides: Mn$_{1-x}$Fe$_x$Ge~\cite{grigoriev2013,shibata2013} and Fe$_{1-x}$Co$_x$Ge~\cite{grigoriev2014}.

 Noteworthy, in case of Mn$_{1-x}$Rh$_x$Ge, the dependence $a(x)$ is a more or less regular convex curve with a maximum at $x=0.5$, while small deviations of both signs in case of Mn$_{1-x}$Co$_x$Ge resemble random varaiations and can be ascribed to uncertainty of experimental data. As mentioned above, a chemical-doping-induced distortion of B20 lattice is largest at $x=0.5$ (see Fig.~\ref{fig:diffraction}). However, substitution of 3d-Mn for 3d-Co, an element of the same period, does not lead to appreciable deviation from Vegard's law, in contrast to substitution for 4d-Rh, which is different from 3d-elements in atomic size, the space distribution of d-orbitals, the depth of potential well. A significant excess of the lattice parameter of Mn$_{0.5}$Rh$_{0.5}$Ge over the linear curve implies that this compound possesses a noticeably looser structure than Mn$_{0.5}$Co$_{0.5}$Ge, which can be explained by a higher degree of disorder in case of 4d-doping.

 Our DFT calculations give an almost linear function $a(x)$ for both Mn$_{1-x}$Co$_x$Ge and Mn$_{1-x}$Rh$_x$Ge (green line in Fig.~\ref{fig:a-x}). The strong deviation from Vegard's law found for Mn$_{1-x}$Rh$_x$Ge~\cite{sidorov2018} is not explained by theory, because we did not take account of structural disorder. However, $a$ is underestimated for MnGe and overestimated for RhGe, so for Mn$_{1-x}$Rh$_x$Ge, slope of the theoretical curve $a(x)$ differs essentially from the experimental one.

 Nevertheless, as a whole, a disagreement between the theoretical and experimental curves does not exceed 0.7\% for Mn$_{1-x}$Co$_x$Ge at all $x$ and for Mn$_{1-x}$Rh$_x$Ge at $x \leq 0.6$. The largest disagreement ($\sim 1.4\%$) in case of RhGe is within uncertainty of measurements and calculations. We tried to change the slope of the theoretical curve for Mn$_{1-x}$Rh$_x$Ge using the GGA$+$U procedure (with different $U$ for Rh and Mn). However, together with improving $a(x)$ with $U$-procedure we always increased at the same time the deviation between experimental and theoretical magnetic moments. Here, it should be noted that DFT explains quite well the evolution of  experimental magnetic moment with concentration $x$, see Fig.~\ref{fig:m-x}.

\begin{figure*}
  \centering
  \includegraphics[width=\columnwidth]{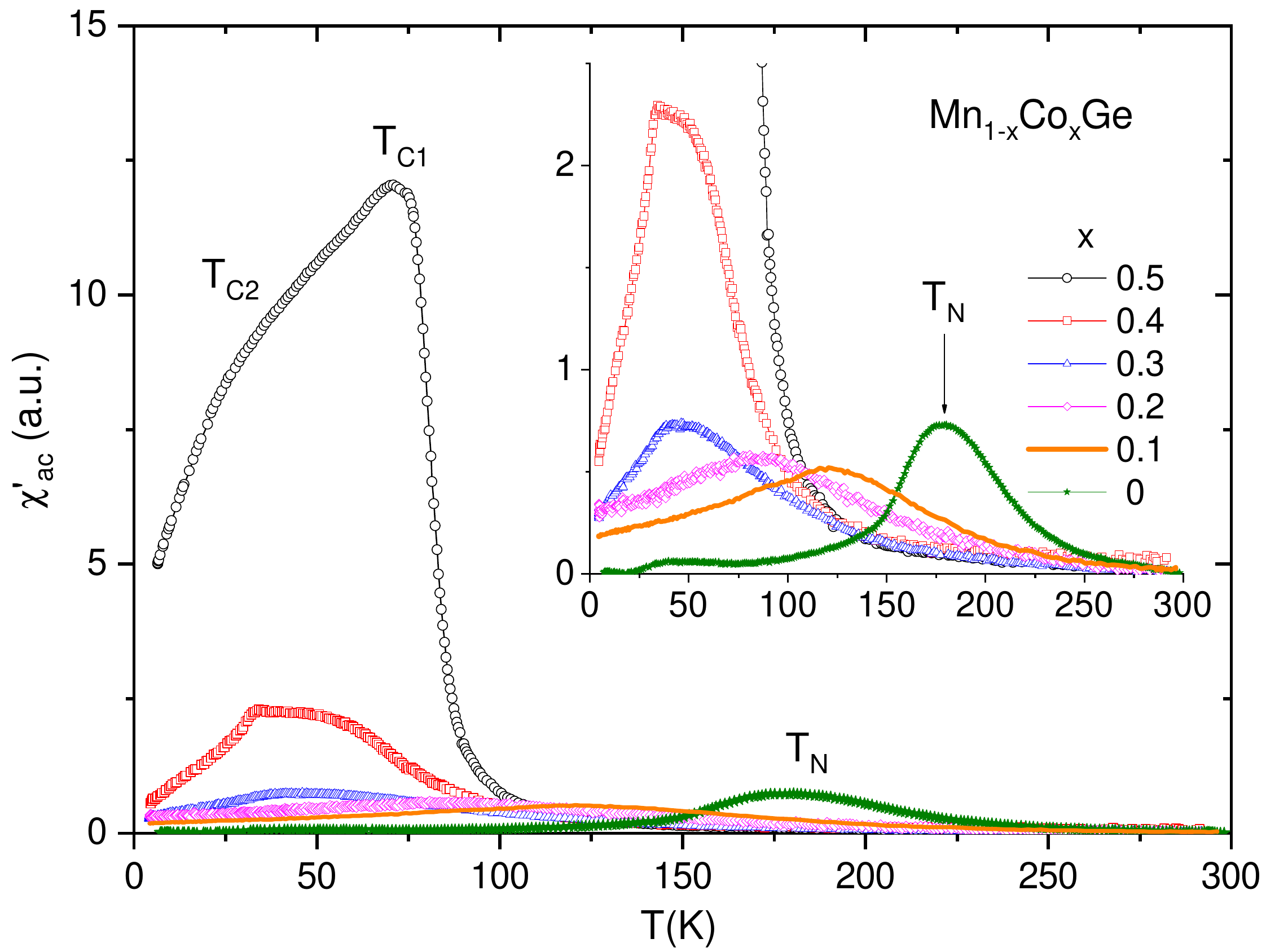}
  \includegraphics[width=\columnwidth]{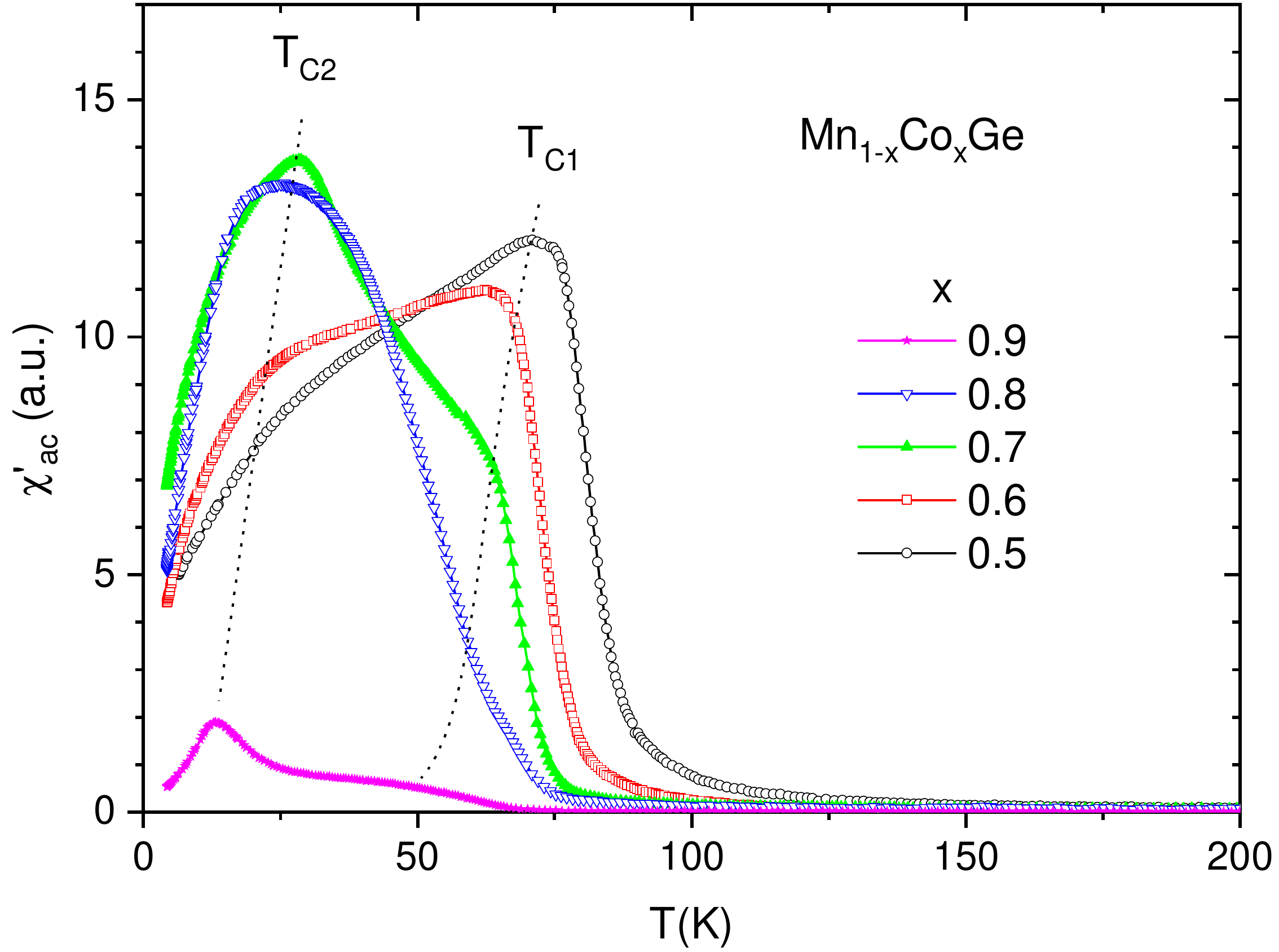}
  \caption{(Color online) The experimental temperature dependence of magnetic susceptibility for Mn$_{1-x}$Co$_x$Ge.}\label{fig:hi}
\end{figure*}

\begin{figure*}
  \centering
  \includegraphics[width=\textwidth]{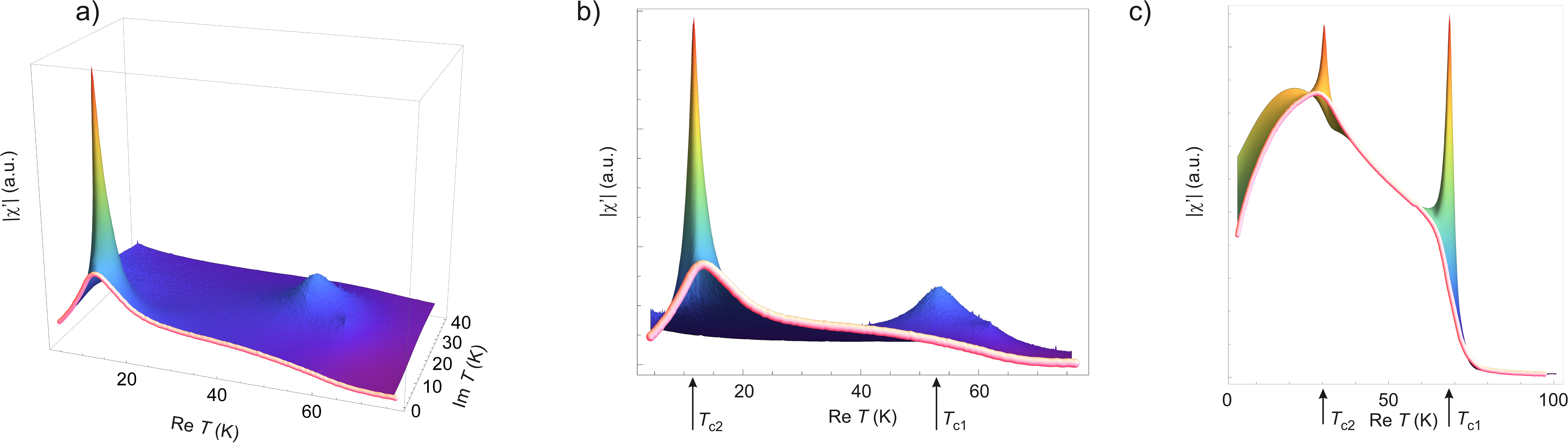}
  \caption{(Color online) Illustrating example of the analytical continuation method applied for  $\chi(T,x=0.9)$ of Mn$_{1-x}$Co$_x$Ge (Fig.~\ref{fig:pade}(a-b)) and $\chi(T,x=0.7)$ (Fig.~\ref{fig:pade}(c)). The thick pink curves are experimental $\chi(T,x)$ (from Fig.~\ref{fig:hi}) interpolated by the Pade-approximant. The surface in Fig.~\ref{fig:pade}(a) is the absolute value of  the analytically continued $\chi(T,x=0.9)$. The front view of  the surface is shown in Fig.~\ref{fig:pade}(b), where the peak positions are now clearly seen.  Fig.~\ref{fig:pade}(c) shows the analytical continuation of $\chi(T)$ for $x=0.7$: two peaks corresponding to the temperatures $T_{c1}$ and $T_{c2}$ of the magnetic transitions are clearly seen.}\label{fig:pade}
\end{figure*}

\section{Magnetic susceptibility of Mn$_{1-x}$Co$_x$Ge and the processing of experimental data\label{Sec4}}

\subsection{Measurement results}
At low temperatures, the compounds Mn$_{1-x}$Co$_x$Ge are found to be magnetically ordered at concentrations of $0 \leq x \leq 0.9$. In order to obtain the magnetic transition temperature ($T_m$) of Mn$_{1-x}$Co$_x$Ge as a function of concentration $x$, we measured the temperature dependencies of the magnetic susceptibility, $\chi(T)$, for some representative concentrations. The obtained susceptibility--temperature curves in the regions $0 \leq x \leq 0.5$ and $0.5 \leq x \leq 0.9$ are displayed respectively in the left and right panels of Fig.~\ref{fig:hi}.

The main feature of the $\chi(T)$ curve is a broad maximum whose position marks temperature where a chiral magnetic ordering emerges. In the Mn-rich side, the $\chi(T)$ curve is similar to that of MnGe. With increasing Co concentration, the main peak is gradually shifted to lower temperatures. The exception is the region of $0.3 \leq x \leq 0.5$, where the peak position goes up. Thus, we found the highest $T_m = 165~K$ for MnGe, the lowest one ($\approx 50~K$) for $ x \approx 0.3$, and a local maximum of $\approx 70~K$ for $ x \approx 0.5$.

As is seen in Fig.~\ref{fig:hi}, the $\chi(T)$ curve at $ x \geq 0.3$ looks such that instead of one broad maximum, two overlapping peaks can be distinguished. This implies that in this concentration range, there are two successive phase transitions to different helimagnetic states (short-period or long-period). We scanned the temperatures of these smeared peaks to prepare the phase diagram (Fig.~\ref{fig:Rh/Co-phase-diag}b) described in Sec.~\ref{Sec5}B.


\subsection{Theoretical procedure for extracting the magnetic transition temperatures}

One of the methods to identify the temperatures of magnetic phase transitions in Mn$_{1-x}$(Co,Rh)$_x$Ge compounds (see Fig.~\ref{fig:Rh/Co-phase-diag}) is related to investigation of the peak positions in the susceptibility curves. However, the peaks are strongly smeared as follows from temperature -- susceptibility curves in Fig.~\ref{fig:hi}. The smearing of the peaks exhibits the appearance of intense helical fluctuations in a wide temperature range. Technically that strongly complicates the definition of magnetic transition temperatures.

We developed the procedure for peak-extraction that removes the ambiguity with the definition of the transition temperature especially when the peaks strongly overlap or hidden behind the helical fluctuations. The idea is to maximally accurately approximate the temperature dependence of the susceptibility using the Pade-approximant, $P_d(T)$, built on top of the experimental data-table. The calculation procedure is described in detail in Refs.~\cite{Chtchelkatchev2015JETPLett,Chtchelkatchev2016JETPLett,Khusnutdinoff2016JETP}.  Then we formally perform the analytic continuation of the susceptibility approximant $P_d(T)$ into the region of complex $T$. Pole-singularities in the complex plain of $T$ help to find the peak location in $\chi(T)$ (with real-valued $T$, of cause). Mathematically, it should be understood that the peak in $\chi(T)$ shows itself as the Lorentzian $L(T)$ (any smooth peak can be in fact approximated like that at least near the peak top):
\begin{equation}
L(T)\propto \frac 1 {(T-T_0)^2+\Gamma^2}\propto \mathrm{Re\,} \frac 1 {T-(T_0+i\Gamma)}.
 \end{equation}
As follows, the Lorentzian in complex plane produces a pole singularity. The width of the Lorentzian, $\Gamma$, (the imaginary part of the pole position) is the estimate of the peak width; the real part -- the peak position.


Building the analytical continuation on top the raw-data is from the first glance quite risky approach that may produce highly unstable results. However it is not so; similar procedure has been recently successfully applied for peak separation of radial distribution function $g(r)$ of liquid alloys where the analytical continuation has been done over the radial coordinate $r$, see Refs.~\cite{Chtchelkatchev2015JETPLett,Chtchelkatchev2016JETPLett,Khusnutdinoff2016JETP}. To ensure the validity of the peak separation procedure, here we use the following trick that helps to avoid instabilities caused by the data inaccuracy: we randomly remove about 10-15$\%$ of data used to build $P_d(T)$ and then average $P_d(T)$ over the samples of data with randomly extracted points. Each random extraction of data may result in some shift of the pole-singularities in $P_d(T)$. After the sample-averaging procedure the pole-singularities smooth in $\langle P_d(T)\rangle$ and look like the ``domes'' in $|\langle P_d(T)\rangle|$ surface in complex plane of $T$. The position of the dome in $|\langle P_d(T)\rangle|$ we attribute to the most probable position of the pole. It should be noted, that this procedure has been tested on a number of toy-functions with well known analytical continuation where we artificially added noise to input data and checked how accurately our method with averaging predicts the position of singularities in the complex plain~\cite{Chtchelkatchev2015JETPLett,Chtchelkatchev2016JETPLett,Khusnutdinoff2016JETP}.

Illustrating example of the analytical continuation method is given in Fig.~\ref{fig:pade} where the method was applied for  $\chi(T,x=0.9)$ of Mn$_{1-x}$Co$_x$Ge (Fig.~\ref{fig:pade}(a-b)) and $\chi(T,x=0.7)$ (Fig.~\ref{fig:pade}(c)). The pink thick curves are experimental $\chi(T,x)$ (from Fig.~\ref{fig:hi}) interpolated by the Pade-approximant. The surface in Fig.~\ref{fig:pade}(a) shows the absolute value of  the analytically continued $\chi(T,x=0.9)$. The front view of  the surface is shown in Fig.~\ref{fig:pade}(b) where the peak-positions are now clearly seen. Similarly,  Fig.~\ref{fig:pade}(c) shows the analytical continuation of $\chi(T)$ for $x=0.7$: two peaks corresponding to the temperatures $T_{c1}$ and $T_{c2}$ of the magnetic transitions are clearly seen.


%

\begin{figure}
  \centering
  \includegraphics[width=\columnwidth]{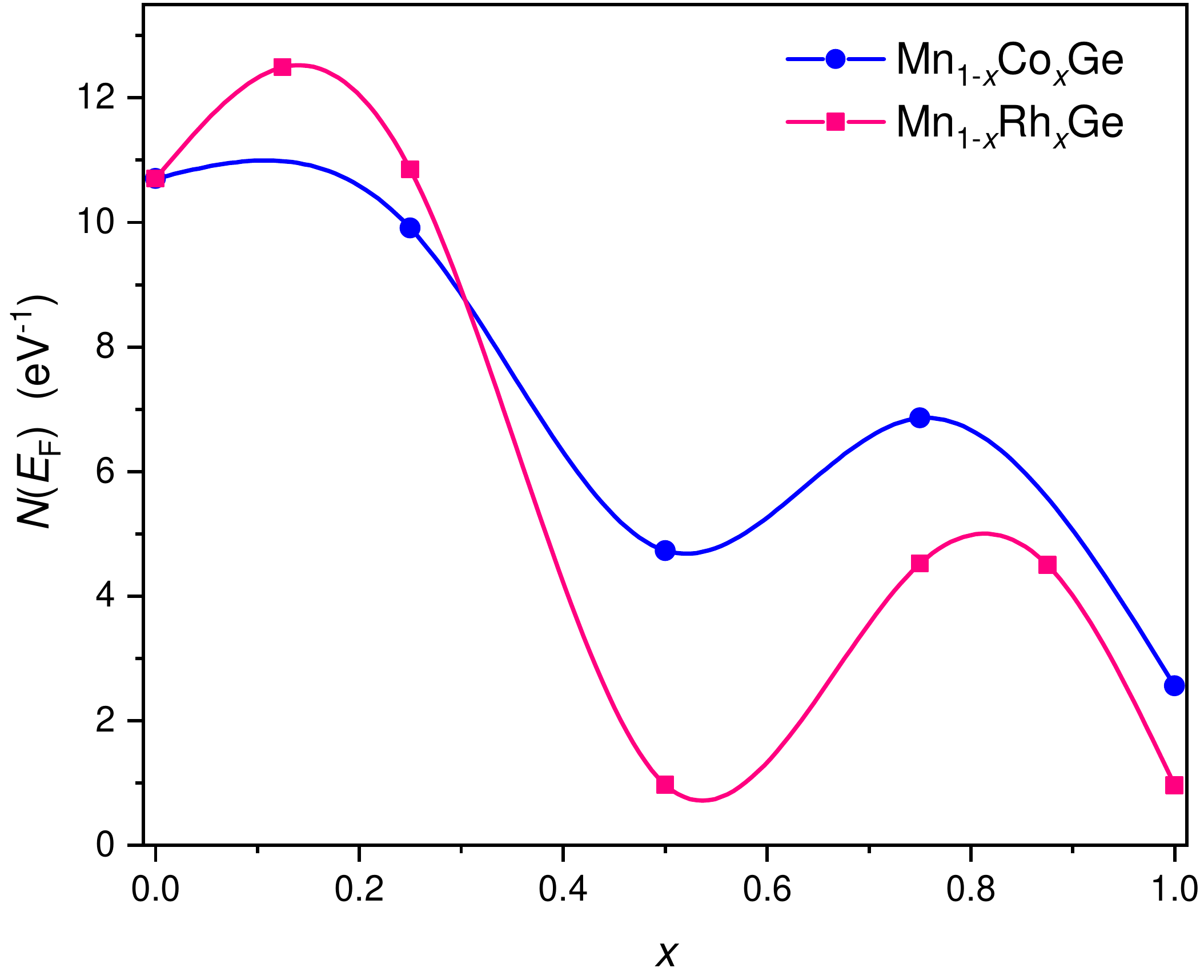}
    \caption{(Color online) Concentration dependence of the density of states at the Fermi level for Mn$_{1-x}$Co$_x$Ge in comparison with Mn$_{1-x}$Rh$_x$Ge~\cite{sidorov2018}.}\label{fig:RhCo-dosEf}

   \includegraphics[width=\columnwidth]{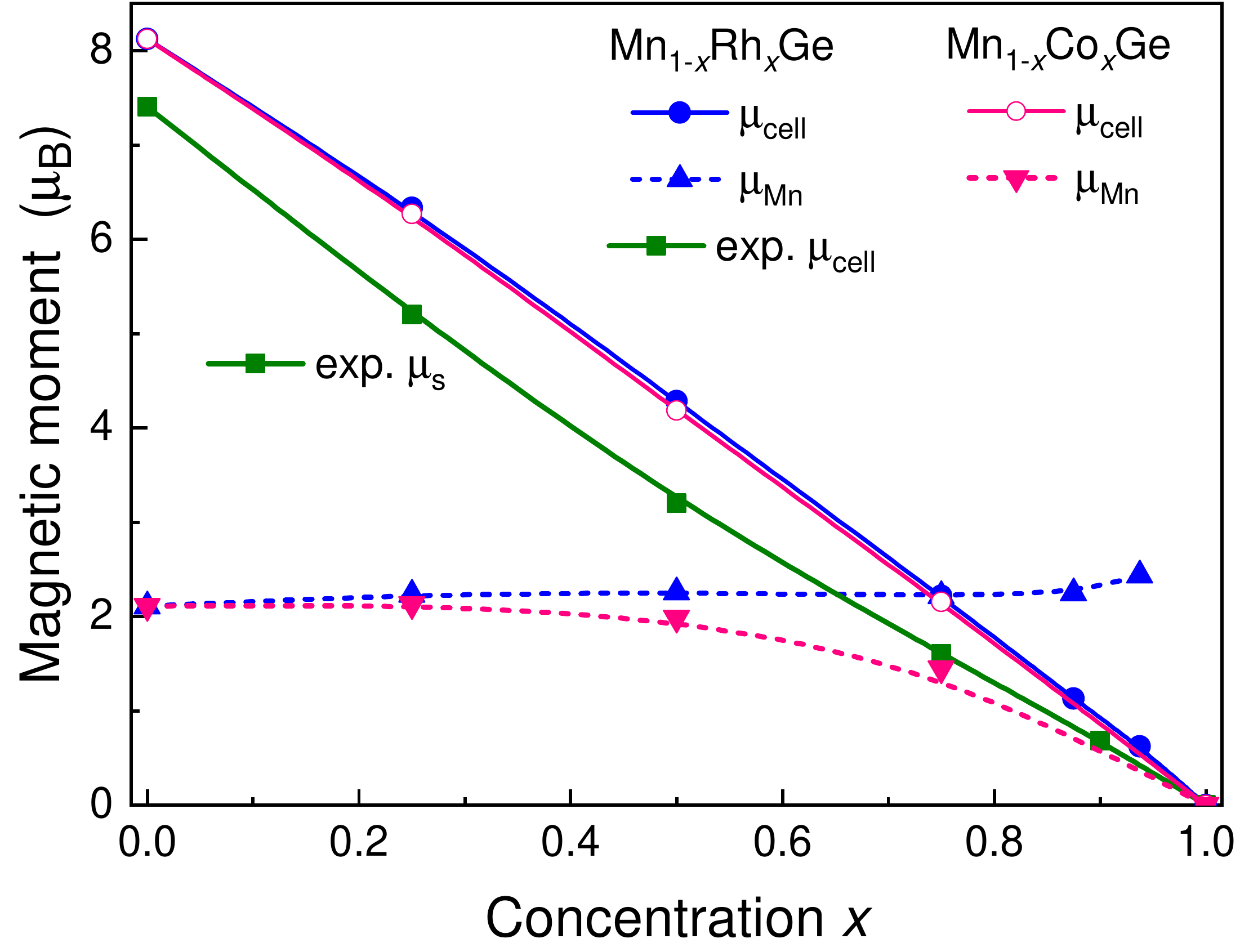}
  \caption{(Color online) Concentration dependence of magnetization per unit cell ($\mu_{\mathrm{cell}}$) and magnetic moment at the Mn atom ($\mu_{\mathrm{Mn}}$) for Mn$_{1-x}$Co$_x$Ge in comparison with the case of Mn$_{1-x}$Rh$_x$Ge~\cite{sidorov2018}. The lines are a guide for the eye.}\label{fig:m-x}
\end{figure}

\section{Comparative discussion of 3d- and 4d-substitution\label{Sec5}}

\subsection{Correlation between the chemical composition and the magnetic properties}


At each particular value of concentration, Mn$_{1-x}$Co$_x$Ge and Mn$_{1-x}$Rh$_x$Ge are isostructural and isovalent compounds. According to the rigid band approximation (RBA), the shape of the bands and their positions relative to the Fermi energy ($E_{\mathrm{F}}$) are analogous for the two compounds. As is seen in Fig.~\ref{fig:RhCo-dos}, the peak positions in the magnetic and nonmagnetic DOS of Mn$_{1-x}$Co$_x$Ge ($x = 0.25, 0.5, 0.75$, and 1) are alike to those of Mn$_{1-x}$Rh$_x$Ge (see Figs. 12 and 13 in~\cite{sidorov2018}).
Correspondingly, the concentration evolution of the DOS at the Fermi level, $N(E_F,x)$, is very similar for both systems (Fig.~\ref{fig:RhCo-dosEf}).

In line with the RBA, the magnetization ($m$) per unit cell should be the same for the both compounds, because it is determined as a difference between the number of spin-up and spin-down electrons. This is the case, which one can see in Fig.~\ref{fig:m-x}, where the dependencies $m(x)$ for Mn$_{1-x}$Co$_x$Ge and Mn$_{1-x}$Rh$_x$Ge practically coincide. Here, the results for Mn$_{1-x}$Rh$_x$Ge are obtained in paper~\cite{sidorov2018}.

The good agreement between the measured and calculated $m(x)$ for Mn$_{1-x}$Rh$_x$Ge demonstrates that the itinerant ferromagnetism model quite well explains the evolution of experimental magnetization with concentration $x$. A small systematic excess of the calculated values over the measured ones can be probably ascribed to the noncollinearity of experimental magnetic arrangement. Unfortunately, there are no magnetization measurements for Mn$_{1-x}$Co$_x$Ge. The available SANS results~\cite{martin2017} obtained for Co/Rh doping up to $ x \leq 0.5$ produce only the magnetic moment at Mn site ($\mu_{\mathrm{Mn}}$) on the assumption that Co/Rh and Ge atoms bear no moment.

Our calculations show that the total magnetization in unit cell is mainly localized at Mn atoms and actually proportional to their number. At $0 \leq x \leq 0.5$, the manganese moment ($\mu_{\mathrm{Mn}} \approx 2 \mu_{\mathrm{B}}$) only slightly depends on concentration $x$. Much smaller moments at Co/Rh atom are parallel to $\mu_{\mathrm{Mn}}$, while a small $\mu_{\mathrm{Ge}}$ (less than $0.1~\mu_{\mathrm{B}}$) is antiparallel to $\mu_{\mathrm{Mn}}$. Fig.~\ref{fig:m-x} displays Mn moment as a function of concentration at $0 \leq x \leq 1$ (the Co/Rh and Ge moments not shown).

It is seen that $\mu_{\mathrm{Mn}}$ decreases/increases upon the Co/Rh substitution. This is clear, because the increasing Co/Rh concentration results in compression/expansion of the lattice. At higher $x$, the values of $\mu_{\mathrm{Mn}}$ for Mn$_{1-x}$Co$_x$Ge and Mn$_{1-x}$Rh$_x$Ge are significantly different, so the equal magnetization per cell is preserved at the expense of different moments at the Co and Rh sites.  For example, at $x \approx 0.94$, $\mu_{\mathrm{Co}}$ and $\mu_{\mathrm{Rh}}$ are equal to 0.38~$\mu_{\mathrm{B}}$ and 0.045~$\mu_{\mathrm{B}}$, correspondingly. Here, it should be noted that the calculation results for the Rh and Ge moments are consistent with the measured sign and $x$-dependence of the XMCD signal for Rh and Ge~\cite{sidorov2018}. Presently, there is no experimental information on the Co and Ge moments in Mn$_{1-x}$Co$_x$Ge.

\subsection{Magnetic phase diagrams}

The magnetic phase diagrams of Mn$_{1-x}$(Co,Rh)$_x$Ge ($0 \leq x \leq 0.5$ ) have been studied using the SANS technique, and for both systems, double long-period structures have been observed at higher doping levels~\cite{martin2017}. The authors~\cite{martin2017} assume that a strong disorder inherent in these high-pressure phases partially destroys the ``ordinary'' spin-spiral structure and stabilizes a state with numerous localized defects, which they call a ``twist grain boundary'' (TGB) phase. The TGB state involves magnetic screw dislocations and is topologically similar to defect networks in smectic liquid crystals. The low-tempearture SANS measurements show that transitions from a simple SPH- to LPH-state occur at $x \approx 0.25$ for Mn$_{1-x}$Rh$_x$Ge and at $x \approx 0.45$ for Mn$_{1-x}$Co$_x$Ge ~\cite{martin2017}.

\subsubsection{ Mn$_{1-x}$Rh$_x$Ge}
\begin{figure}
  \centering
   \includegraphics[width=\columnwidth]{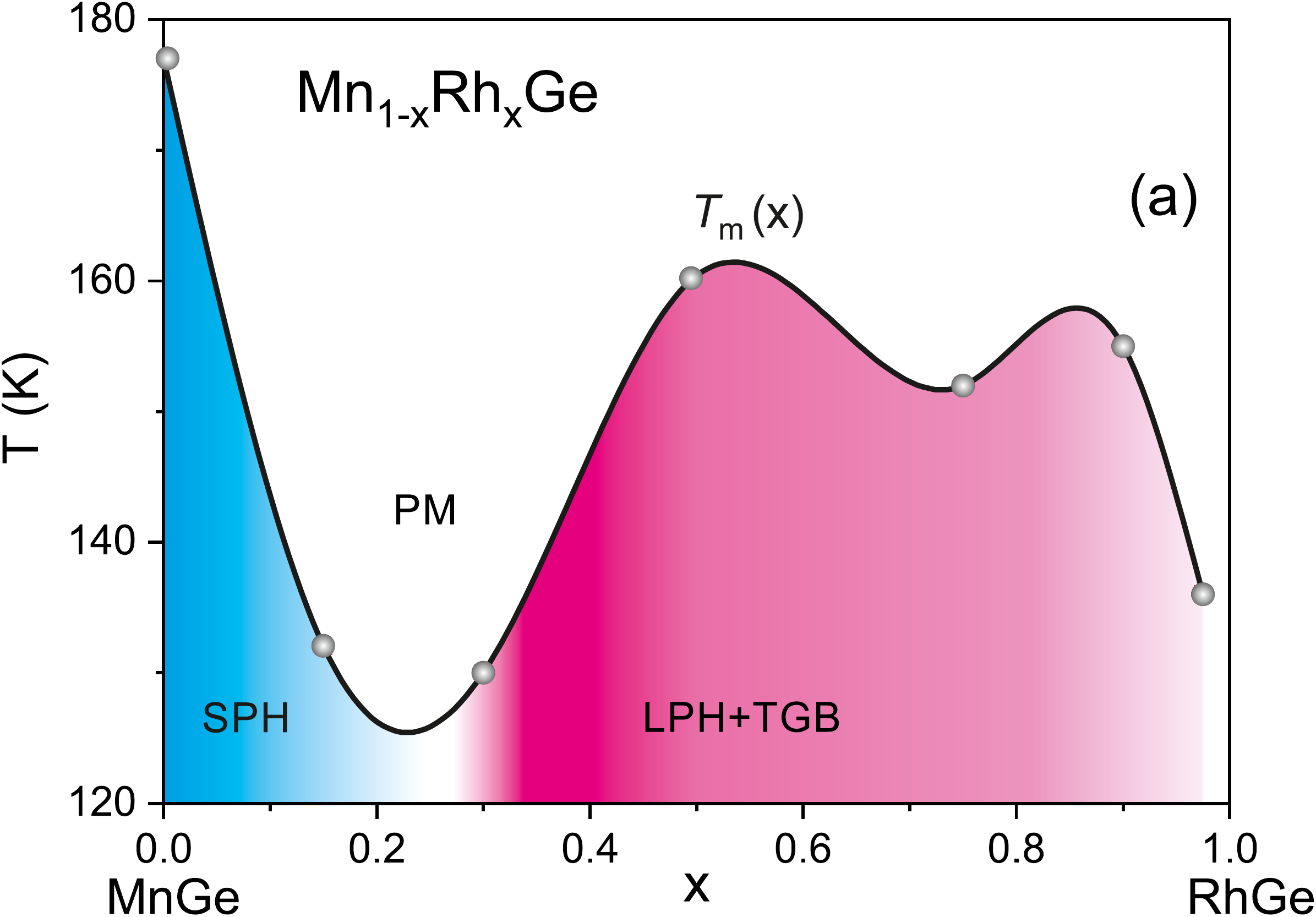}
 \\
  \includegraphics[width=\columnwidth]{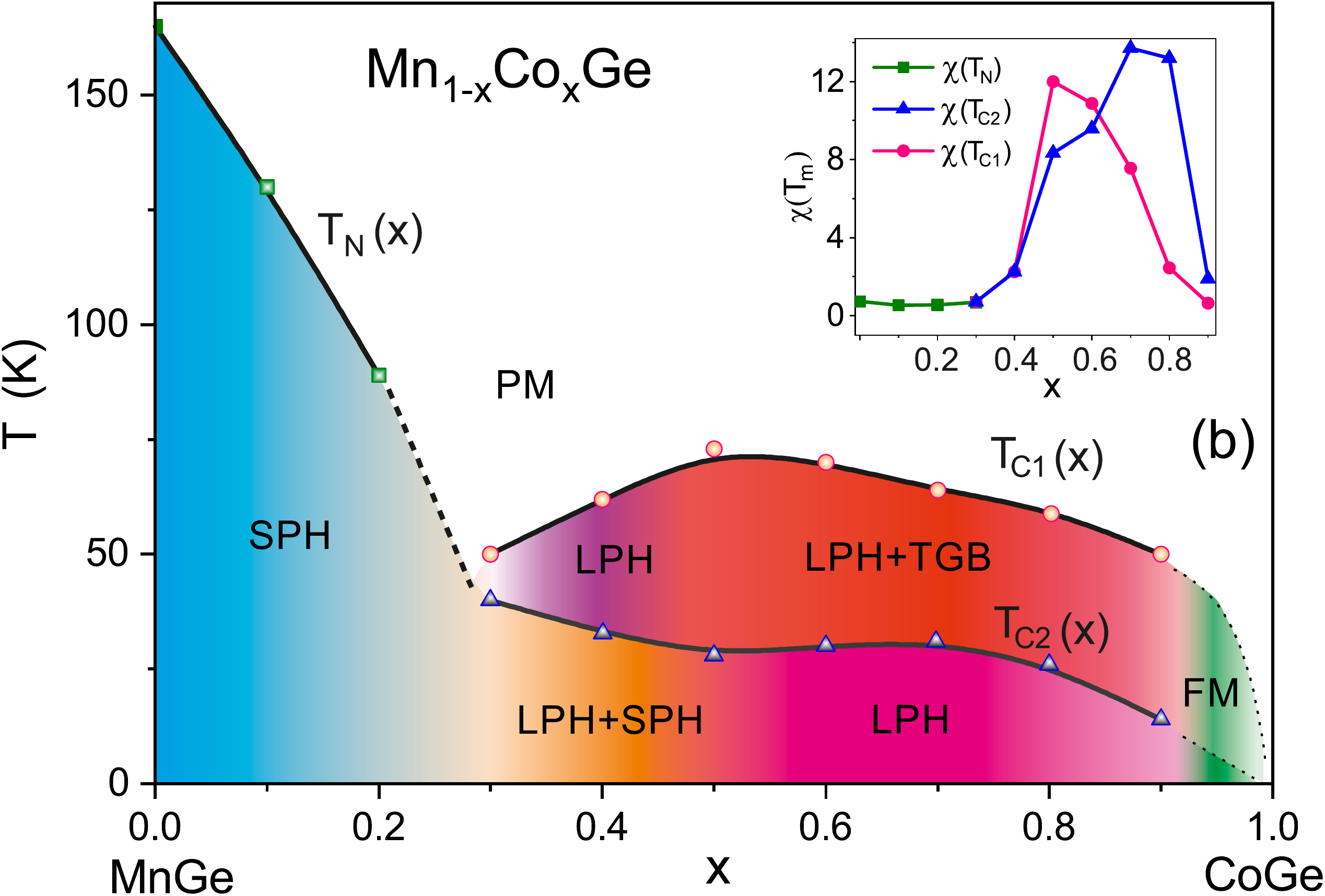}
 \caption{(Color online) $T$--$x$ phase diagram of the magnetic structure of Mn$_{1-x}$Rh$_x$Ge (a) and Mn$_{1-x}$Co$_x$Ge (b). The inset in the figure (b) shows the magnetic susceptibility value at transition temperatures.}\label{fig:Rh/Co-phase-diag}
\end{figure}

Figure~\ref{fig:Rh/Co-phase-diag}a displays the $T$--$x$ magnetic phase diagram for Mn$_{1-x}$Rh$_x$Ge constructed using the results of our previous susceptibility measurements at $0 \leq x \leq 1$~\cite{sidorov2018}, with taking account of the SANS results at $0 \leq x \leq 0.5$~\cite{martin2017}. Two main regions of dissimilar magnetic behavior are filled with different colours. To be more specific, between $x = 0.2$ and 0.3, a change occurs from a SPH ($\sim 30$~{\AA} as in pure MnGe) to a combination of a LPH ($\sim 380$~{\AA}) and a TGB phase. According to paper ~\cite{martin2017}, the LPH+TGB state exists in the composition range $0.25 \leq x \leq 0.5$.

Our measurements~\cite{sidorov2018} demonstrate similar temperature and pressure behavior of magnetic susceptibility over the range $0.5 \leq x \leq 0.975$. In this region, the magnetic transition temperature $T_m$ only slowly changes with increasing $x$ and increases with increasing pressure ($p$), while in the Mn-rich end, the dependence $T_m(p)$ decreases. So we suppose that the LPH+TGB state is preserved above $x = 0.5$ and up to $x = 0.975$ (this region in Fig.~\ref{fig:Rh/Co-phase-diag}a is painted over with light pink color). This assumption remains to be checked in the future SANS experiments.

A strikingly nonmonotonic concentration dependence of magnetic transition temperature, $T_m(x)$, observed in Fig.~\ref{fig:Rh/Co-phase-diag} might be qualitatively explained as follows. In paper~\cite{Szpunar1977PSS}, a very simple model for alloys with localized magnetic moments has been proposed, which assumes interaction of the RKKY type mediated by the conduction electrons of the alloy. This approach relates the magnetic transition temperature $T_m$ and the dopant concentration:

\begin{equation}\label{eq:Tc}
k_B T_m=\frac 13 N_{\rm alloy}(E_F)\sum_a S_a(S_a+1) j_{a}^2 c_a.
\end{equation}

Herein, $a$ labels the type of component, $c_a$ and $S_a$ are respectively the concentration and total spin of the component, $j_{a}$ is the effective exchange matrix element, specific to the magnetic component atom and nearly independent of its surroundings, and $N_{\rm alloy}(E_F)$ is the concentration-weighted average alloy density of states at $E_F$. Equation~(\ref{eq:Tc}) is quite accurate, e.g., it has allowed to correctly reproduce the experimental dependence of $T_m(x)$ in the alloys Gd(Al$_{1-x}$M$_x$)$_2$~\cite{Magnitskaya1994PRB}.

Figuratively, the relation like Eq.~(\ref{eq:Tc}) should be applicable also for the 3d Mn-based alloys in hand. If so, $k_B T_m\propto (1-x) N(E_F,x)$, which suggests a correlation between the experimental transition temperature $T_m(x)$ of, e.g., Mn$_{1-x}$Rh$_x$Ge (Fig.~\ref{fig:Rh/Co-phase-diag}a), and the nonmonotonic curve $N(E_F,x)$ in Fig.~\ref{fig:RhCo-dosEf}(red line).

\subsubsection{ Mn$_{1-x}$Co$_x$Ge}
Compared to the case of Rh substitution, the magnetic phase diagram of Mn$_{1-x}$Co$_x$Ge is expected to be more complicated, due to combination of two magnetic 3d-elements (4d-Rh is paramagnetic). Actually, it represents a complex sequence of different magnetic orders and crossovers (Fig.~\ref{fig:Rh/Co-phase-diag}b). The analysis of Sec.~\ref{Sec4} allowed us to reliably resolve the double-peak features of experimental $\chi(T)$ dependencies and produce this $T-x$ phase diagram. Here, it should be mentioned that the results of our high-sensitive measurements of $\chi(T)$ confirm and supplement the data~\cite{altynbaev2018} obtained with the SQUID magnetometer. The SANS results~\cite{martin2017,altynbaev2018} are also taken into account.

As is seen in Fig.~\ref{fig:Rh/Co-phase-diag}, instead of two composition ranges of different magnetic behavior distinguished for Rh substitution, we observe four such ranges for Mn$_{1-x}$Co$_x$Ge. Another sharp distinction from Mn$_{1-x}$Rh$_x$Ge is the presence of two temperature intervals, whose upper borders are denoted as $T_{C1}(x)$ and $T_{C2}(x)$.

It is seen in Fig.~\ref{fig:Rh/Co-phase-diag}b that in the Mn-rich side ($x < 0.25$), a SPH with a period of $\sim 30$~{\AA} exists in a wide range of temperatures up to 165~K (painted blue in the figure). Next interval of concentrations is $0.25 < x < 0.5$, where a low-temperature SPH coexists with a LPH up to $T_{C2} \sim 40~K$. Within this region, the helix period increases with $x$ up to $\sim 120$~{\AA} (see paper~\cite{altynbaev2018}, where Fig. 10 displays the $x$-dependence of the helix wave vector \textbf{Q} measured at $T = 5~K$). For the coexisting LPH, the temperature of transition to paramagnetic state $T_{C1} > T_{C2}$, so the LPH extends up to $T_{C1} \sim 70~K$. With regard to Co concentration, this LPH extends up to $x = 0.85$. The 5~K helix period continues to increase with increasing concentration and at $ x\sim 0.8$ reaches values on the order of $\sim 550$~{\AA}. Further increase in concentration $x$ results in the stabilization  at $x > 0.9$ of ferromagnetic-like ordering (green region) -- so to speak, the limit of extremely long spiral -- and nonmagnetic state for pure CoGe.

The region [$0.5 < x < 0.85$; $T_{C2} < T < T_{C1}$] is of special interest: although the SPH does not exist there, the double-peak shape of $\chi(T)$ dependencies is preserved. In paper~\cite{altynbaev2018}, the analysis of the experimental data for two compositions, Mn$_{0.4}$Co$_{0.6}$Ge and Mn$_{0.5}$Co$_{0.5}$Ge, has also demonstrated the presence of two-peak feature of SANS intensity in the temperature range from 5~K to 80~K. This feature has been described in~\cite{altynbaev2018} as consisting of main and satellite peaks. The authors~\cite{martin2017}, however, explain the second peak as a manifestation of the TGB phase. We join the latter opinion and consider the magnetic structure of Mn$_{1-x}$Co$_x$Ge in this region as the  LPH+TGB coexistence.

Thus, the entire domain where the LPH is observed covers the concentrations from 0.25 to about 0.9 and temperatures below $T_{C1}(\sim70$~K). At lower Co concentrations ($0.25 < x < 0.5$) and below $T_{C2}$ ($\sim40$~K), it coexists with the low-temperature SPH. Then, at $0.5 < x < 0.85$, the LPH coexists with the high-temperature TGB state whose temperature region of stability is between $T_{C2}(x)$ and $T_{C1}(x)$.

\begin{figure}
  \centering
  \includegraphics[width=\columnwidth]{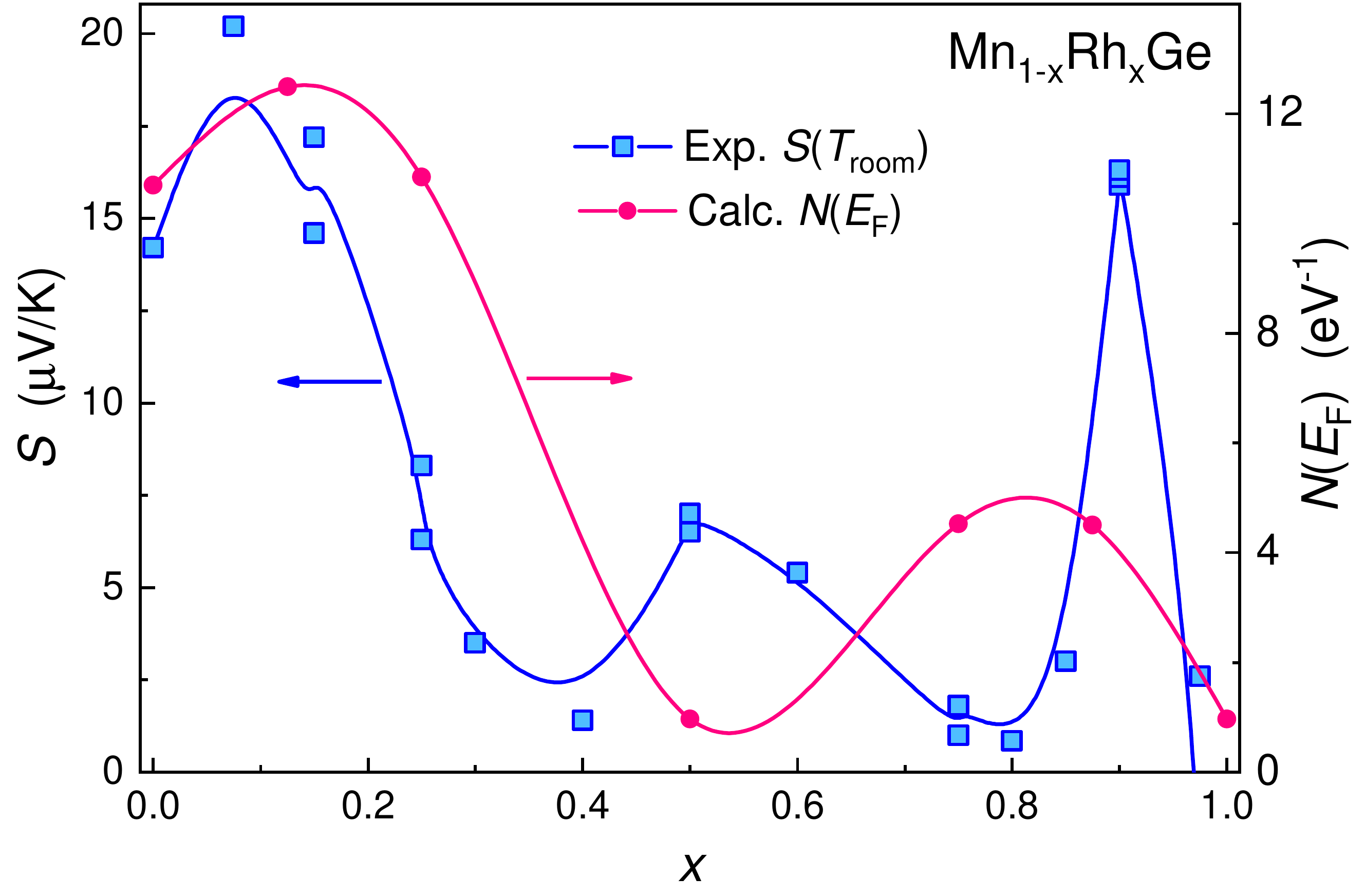}
  \caption{(Color online) Concentration dependencies of the room-temperature Seebeck coefficient $S$ and density of states at the Fermi level $N(E_F)$ for Mn$_{1-x}$Rh$_x$Ge.}\label{fig:Rh-S-dosEf}
\end{figure}


\subsection{Transport properties}

\begin{figure*}
  \centering
  \includegraphics[width=\columnwidth]{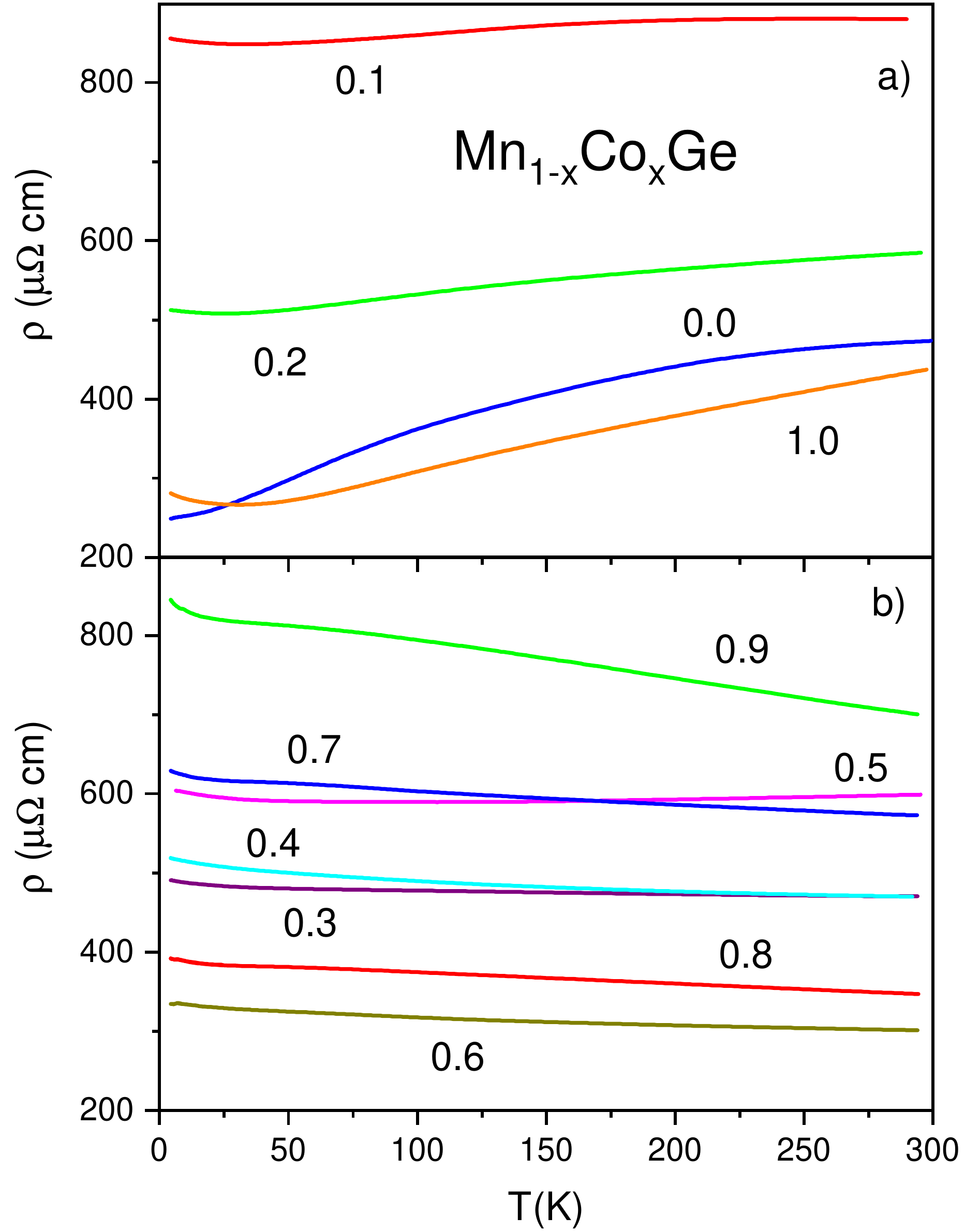}
  \includegraphics[width=\columnwidth]{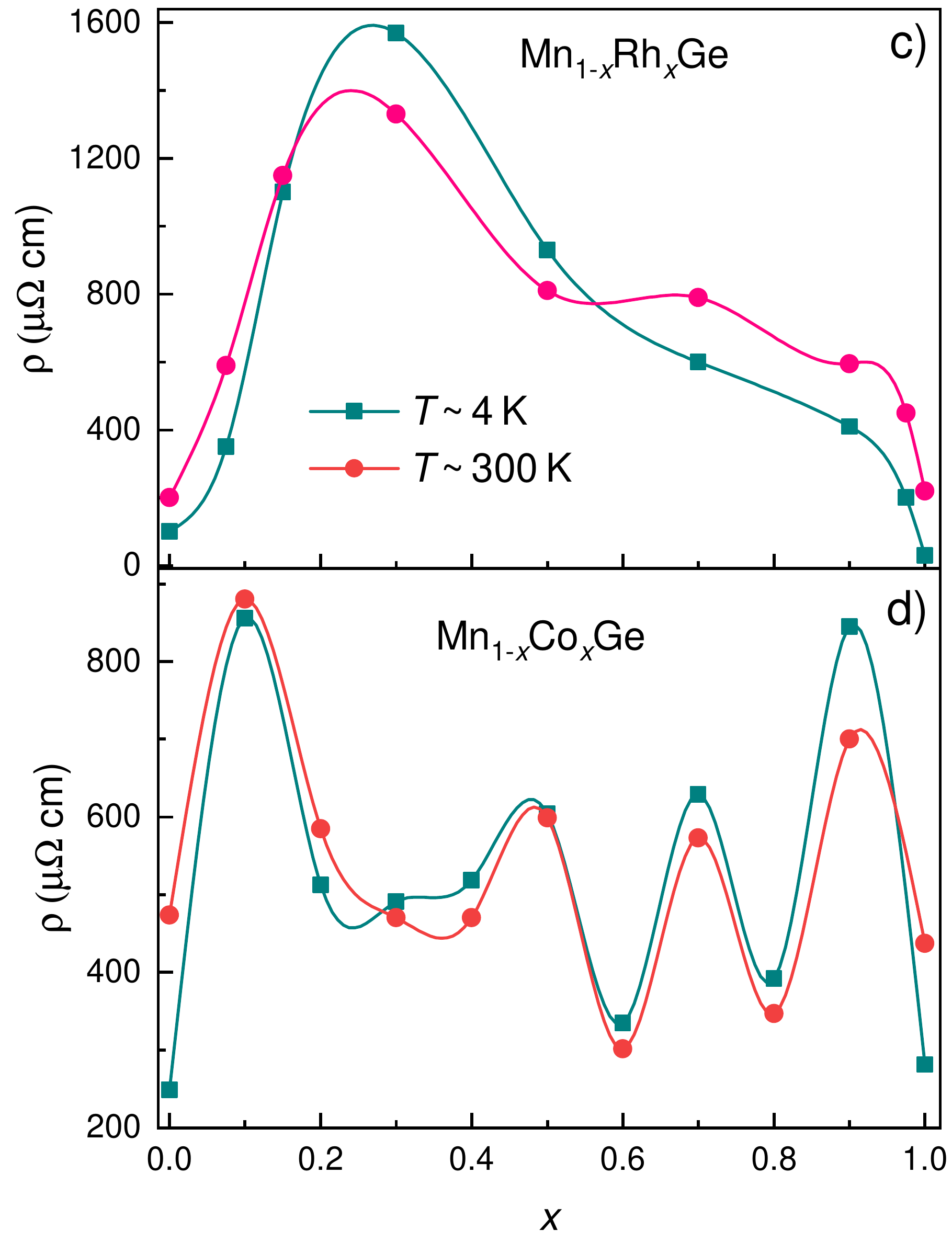}
  \caption{(Color online) Left, Temperature dependence of electrical resistivity in Mn$_{1-x}$Co$_x$Ge at low $x$ and $x = 1$ (a) and at intermediate $x$ (b). Right, Concentration dependence of the low- and room-temperature resistivity for Mn$_{1-x}$Rh$_x$Ge (c) and Mn$_{1-x}$Co$_x$Ge (d).}\label{fig:rho-x}
\end{figure*}

The  transport properties of Mn$_{1-x}$Co$_x$Ge show unusual behavior on going from $x = 0$ to 1. Figure~\ref{fig:Rh-S-dosEf} displays the $x$-dependencies of the experimental Seebeck coefficient~\cite{sidorov2018} and the DOS at the Fermi level for Mn$_{1-x}$Rh$_x$Ge. Here, we present only a positive part of the curve $S(x)$ (Seebeck coefficient of the Rh-rich end is negative). We observe a rough correlation between the nonmonotonic curves $S(x)$ and $N(E_{\mathrm{F}},x)$  with respect to extrema positions.

The electrical resistivity $\rho$ of Mn$_{1-x}$Co$_x$Ge and its temperature dependence $\rho(T)$ also strongly change with $x$. The end members, MnGe and CoGe, are metal-like conductors with residual resistivity $\rho_{\mathrm{res}}\sim 250 \mu \Omega$~cm and room-temperature resistivity $\rho_{\mathrm{300K}}\sim 450 \mu \Omega$~cm. As is seen in Fig.~\ref{fig:rho-x}a, for CoGe, MnGe, and two Mn-rich compositions $x = 0.1$ and 0.2, the temperature dependence of resistivity is similar to that of metals. In contrast, at intermediate $x$ values, there appears a ``semiconducting'' behavior of $\rho(T)$ (Fig.~\ref{fig:rho-x}b) quite typical of disordered and partially ordered alloys, see, e.g., Refs.~\cite{Gunnarsson2003RMP,Uporov2018JAC}.

Fig.~\ref{fig:rho-x} (right) shows the $x$-dependence of resistivity for Mn$_{1-x}$(Co,Rh)$_x$Ge. Contrary to the case of Mn$_{1-x}$Rh$_x$Ge, the dependence $\rho(x)$ for Mn$_{1-x}$Co$_x$Ge exhibits strong oscillations, which seemingly reflects the more complicated magnetic phase diagram of the latter. In the case of Rh substitution, the ``semiconducting'' behavior, $\rho_{\mathrm{res}} > \rho_{300\mathrm{K}}$, is observed in the range $0.2 \lesssim x \lesssim 0.5$ (Fig.~\ref{fig:rho-x}c), where the existence of LPH is proved in SANS experiments. For Co substitution, $\rho_{\mathrm{res}}$ exceeds $\rho_{300\mathrm{K}}$ between $x\simeq$ 0.25 and $x\simeq$ 0.95 (Fig.~\ref{fig:rho-x}d), i.e. also in the domain of the LPH existence (see Fig.~\ref{fig:Rh/Co-phase-diag}). This can be accounted for by the conjecture that magnetic fluctuations contribute to electron scattering~\cite{Kataoka2001PRB} in the transitive region between the SPH behavior in the MnGe-rich end and the ferro/paramagnetic state in the (Co,Rh)-rich end.

In the present context, two remarks can be added: Even in the regions of metallic behavior of $\rho(T)$, the residual resistivity $\rho_{\rm res}$ of Mn$_{1-x}$(Co,Rh)$_x$Ge substantially exceeds the temperature-dependent contribution $\rho_{300} - \rho_{\rm res}$, which suggests a large number of crystallographic defects in polycrystalline samples of these metastable high-pressure phases. Lastly, a difference in the resistivity values for pure MnGe obtained in two batches of measurements (Figs.~\ref{fig:rho-x}c and ~\ref{fig:rho-x}d) is related to the quality of experimental samples.


\section{Conclusions\label{Sec6}}
To conclude, we performed experimental and theoretical study of the high-pressure-synthesized chiral magnets Mn$_{1-x}$Co$_x$Ge ($0 \leq x \leq 1$). The high-sensitive experimental technique was applied to measure the magnetic ac-susceptibilities $\chi$ from 4~K to 300~K. The use of low excitation field of about 1 Oe allowed us to resolve the double-peak shape of $\chi(T)$ smeared out in higher fields and thereby provide a new $T$--$x$ magnetic phase diagram of Mn$_{1-x}$Co$_x$Ge.

Our experimental data were theoretically analyzed on the basis of  \textit{ab initio} DFT calculations of Mn$_{1-x}$Co$_x$Ge. The results of measurements and calculations were compared with the data on Mn$_{1-x}$Rh$_x$Ge obtained in our previous study~\cite{sidorov2018} and reprocessed and summarized in the present paper. This allowed us to trace the effect of progressive 3d/4d-substitution on the structural, electronic, magnetic, and transport properties of the two continuous series of solid solutions Mn$_{1-x}$(Co,Rh)$_x$Ge.

The results of DFT computations were processed using the VESTA software package and the resulting simulated XRD patterns were compared with our experimental powder-diffraction data for MnGe, RhGe, CoGe, and Mn$_{0.5}$Co$_{0.5}$Ge. As follows, at intermediate concentrations $0 < x < 1$, a symmetry-breaking distortion of the B20 lattice takes place. So the pseudobinary compounds can be only approximately considered as B20-type (the experimental B20 reflections in the pseudobinaries are slightly broadened). However, even in the case of maximal distortion, Mn$_{0.5}$(Co,Rh)$_{0.5}$Ge, deviation from the B20 symmetry is not large.

We also show that the experimentally measured concentration dependence of the lattice constant $a(x)$ for Mn$_{1-x}$Co$_x$Ge is practically linear, i.e. obeys Vegard's law. This observation confirms the available structural studies of Mn$_{1-x}$Co$_x$Ge and is consistent with the results for other B20-type 3d-metal monogermanides.

For Mn$_{1-x}$Rh$_x$Ge, however, a strong positive deviation from Vegard's law was found, namely the dependence $a(x)$ is a convex curve with a maximum at $x = 0.5$. A significant excess of $a(x)$ over the linear curve for Mn$_{0.5}$Rh$_{0.5}$Ge suggests a noticeably looser structure than in case of Mn$_{0.5}$Co$_{0.5}$Ge. This is a consequence of higher lattice disorder in case of 4d-substitution, because the atomic radius of 4d-Rh is larger than that of 3d elements. In addition, compared to 3d, the 4d elements are characterized by deeper potential wells, more even space distribution of d-orbitals, etc.

 Our calculated $a(x)$ is almost linear for both systems Mn$_{1-x}$(Co,Rh)$_x$Ge. The positive deviation from linearity in case of Rh-substitution is not reproduced by theory, probably, because we did not take account of structural disorder. Anyway, the largest disagreement between the theoretical and experimental $a(x)$ ($\sim 1.4\%$ for RhGe) is within uncertainty of measurements and calculations.

We found a good agreement between the measured and calculated magnetization $m(x)$ for Mn$_{1-x}$Rh$_x$Ge, which demonstrates that the itinerant ferromagnetism model quite well explains the evolution of experimental magnetization with concentration $x$. According to calculations, the total magnetization in unit cell is mainly localized at Mn atoms and actually proportional to their number, with the manganese moment $\mu_{\mathrm{Mn}}\sim2~\mu_{\mathrm{B}}$. Much smaller moments at Co/Rh atoms are parallel to $\mu_{\mathrm{Mn}}$, while the small Ge moment ($\mu_{\mathrm{Ge}} < 0.1~\mu_{\mathrm{B}}$) is antiparallel to $\mu_{\mathrm{Mn}}$. The results for the Rh and Ge moments are consistent with the measured sign and $x$-dependence of the XMCD signal for Rh and Ge. Presently, there is no experimental information on the total magnetization per cell, $\mu_{\mathrm{Co}}$, and $\mu_{\mathrm{Ge}}$  in Mn$_{1-x}$Co$_x$Ge.

Our measured $\chi(T)$ curves for Mn$_{1-x}$Co$_x$Ge look such that at $ x\geq0.3$, instead of one broad maximum, two overlapping peaks can be distinguished. This implies that at these concentrations,  there are two successive phase transitions to different helimagnetic states (short-period or long-period).  We developed theoretical procedure for extracting the magnetic transition temperatures from $\chi(T)$. It is based on the Pade approximation of $\chi(T)$ and the analytical continuation into complex $T$ plane. It has been shown that the study of pole-singularities allows to extract the peak position -- transition temperature -- even when the peaks strongly overlap and  are ``hidden'' in $\chi(T)$.

Based on $\chi(T)$ measurements, with regard to available SANS data, we provide the new magnetic phase diagram of Mn$_{1-x}$Co$_x$Ge and compare it with that of Mn$_{1-x}$Rh$_x$Ge constructed using our previous experimental data. The both phase diagrams are consistent in general outline: they are characterized by a minimum of $T_m(x)$ between $x = 0.2$ and 0.3, where the fraction of substituting atoms becomes more appreciable. In this region, a SPH-to-LPH crossover takes place. Symmetrically, there is a dip of $T_m(x)$ in the Rh-rich end in Mn$_{1-x}$Rh$_x$Ge ($x \simeq 0.75$). For both systems, $T_m(x)$ has a maximum around $x = 0.5$, which corresponds to the highest lattice disorder. According to the SANS data, near the composition Mn$_{0.5}$Co$_{0.5}$Ge, a change occurs from LPH+SPH to LPH+TGB state. Then, a narrow range of ferromagnetic-like behavior is observed for Mn$_{1-x}$Co$_x$Ge above $x = 0.9$, and finally, both pure RhGe and CoGe are nonmagnetic.

Thus, the nonmonotonic behavior of $T_m(x)$ is analogous for both systems and roughly correlates with the $x$-dependence of the density of states at the Fermi level, $N(E_{\mathrm{F}},x)$, with respect to extrema positions. Qualitatively similar correlation is also observed for the $x$-dependence of Seebeck coefficient $S(x)$. As a matter of fact, the function $N(E_{\mathrm{F}},x)$ gives the number of d-electrons near the Fermi energy, which defines electronic and magnetic properties at particular concentration $x$.

It should be also noted that the electrical resistivity for Mn$_{1-x}$(Co,Rh)$_x$Ge exhibits the ``semiconducting'' behavior of $\rho(T)$ at $0.25 \lesssim x \lesssim 0.5$. It is a transitive region from the SPH to LPH order, where the helix period as a function of $x$ rises most steeply.

In general, the concentration $x$ of substituting element determines such characteristics of the system as symmetry, lattice parameters, structural disorder, magnetic order, transport properties, etc. Actually, at each particular value of 3d-Co (4d-Rh) concentration $x$, the compounds Mn$_{1-x}$(Co,Rh)$_x$Ge are isostructural and isovalent, with their ground-state properties well described within the rigid band approximation. The Vegard's and non-Vegard's behavior of lattice constant $a(x)$ on Co- and Rh-substitution, respectively, can be explained by the larger atomic size of 4d-Rh, which leads to a larger degree of lattice disorder at intermediate concentrations. Another noticeable difference is the more complicated magnetic phase diagram of Mn$_{1-x}$Co$_x$Ge related to the fact that, unlike rhodium, cobalt  is magnetic 3d-element.

\begin{acknowledgments}
The authors gratefully thank S.~M.~Stishov for interest to this work and acknowledge valuable discussions with Yu.~A.~ Uspenskii and I.~Mirebeau. This work was supported by Russian Science Foundation: N.M.C. and M.V.M. acknowledge the support of their theoretical calculations (grant RSF 18-12-00438); V.A.S., A.E.P., and A.V.T. are grateful for support of their experimental measurements (grant RSF 17-12-01050). The numerical calculations were carried out using computing resources of the federal collective usage center `Complex for Simulation and Data Processing for Mega-science Facilities` at NRC `Kurchatov Institute` (http://ckp.nrcki.ru/) and supercomputers at Joint Supercomputer Center of Russian Academy of Sciences (http://www.jscc.ru). We also thank for access to the URAN cluster (http://parallel.uran.ru) made by the Ural Branch of Russian Academy of Sciences.
\end{acknowledgments}

\bibliography{OURbib}
\end{document}